\documentclass[10pt,twocolumn,letterpaper]{article}
\usepackage[accsupp]{axessibility}
\usepackage{appendix}
\usepackage[pagenumbers]{cvpr} 

\usepackage[dvipsnames]{xcolor}

\definecolor{Gray}{gray}{0.9}

\newcommand{\blur}{\boldsymbol{\mathcal{B}}}
\newcommand{\tilt}{\boldsymbol{\mathcal{T}}}
\newcommand{\warp}{\boldsymbol{\mathcal{W}}}

\usepackage{commands}
\usepackage{color, colortbl}
\usepackage{xcolor}
\usepackage{pifont}
\usepackage{subfloat}
\usepackage{booktabs}
\usepackage{diagbox}
\usepackage{multirow}
\usepackage{makecell}
\usepackage{amsmath}
\usepackage{algorithm}
\usepackage{algpseudocode}
\usepackage{cuted}

\makeatletter
\algnewcommand{\LineComment}[1]{\Statex \hskip\ALG@thistlm \(\triangleright\) #1}
\makeatother
\definecolor{cvprblue}{rgb}{0.21,0.49,0.74}
\usepackage[pagebackref,breaklinks,colorlinks,citecolor=cvprblue]{hyperref}

\def\paperID{13203} 
\def\confName{CVPR}
\def\confYear{2024}

\title{Spatio-Temporal Turbulence Mitigation: A Translational Perspective}

\author{Xingguang Zhang$^1$ \ \ \  Nicholas Chimitt$^1$ \ \ \ Yiheng Chi$^1$ \ \ \ Zhiyuan Mao$^2$ \ \ \ Stanley H. Chan$^1$\\
  $^1$School of Electrical and Computer Engineering, Purdue University
  $^2$Samsung Research America \\
  {\tt\small \{zhan3275, nchimitt, chi14, stanchan\}@purdue.edu, m940421@gmail.com}
}

\begin{document}
\maketitle

\begin{abstract}
Recovering images distorted by atmospheric turbulence is a challenging inverse problem due to the stochastic nature of turbulence. Although numerous turbulence mitigation (TM) algorithms have been proposed, their efficiency and generalization to real-world dynamic scenarios remain severely limited. Building upon the intuitions of classical TM algorithms, we present the Deep Atmospheric TUrbulence Mitigation network (DATUM). DATUM aims to overcome major challenges when transitioning from classical to deep learning approaches. By carefully integrating the merits of classical multi-frame TM methods into a deep network structure, we demonstrate that DATUM can efficiently perform long-range temporal aggregation using a recurrent fashion, while deformable attention and temporal-channel attention seamlessly facilitate pixel registration and lucky imaging. With additional supervision, tilt and blur degradation can be jointly mitigated. These inductive biases empower DATUM to significantly outperform existing methods while delivering a tenfold increase in processing speed. A large-scale training dataset, ATSyn, is presented as a co-invention to enable the generalization to real turbulence. Our code and datasets are available at \href{https://xg416.github.io/DATUM}{\textcolor{pink}{https://xg416.github.io/DATUM}}
\end{abstract}    
\section{Introduction}
\label{sec:intro}

Atmospheric turbulence is a dominant image degradation for long-range imaging systems. Reconstructing images distorted by atmospheric turbulence is an important task for many civilian and military applications.
The degradation process can be considered a combination of content-invariant random pixel displacement (i.e., tilt) and random blur.
Until recently, reconstruction algorithms have often been in the form of model-based solutions, often relying on modalities such as pixel registration and deblurring.
Although there have been many important insights into the problem, e.g., lucky imaging, they are primarily limited to static scenes with slow processing speed. 


\begin{figure}[t]
  \centering
   \includegraphics[width=\linewidth]{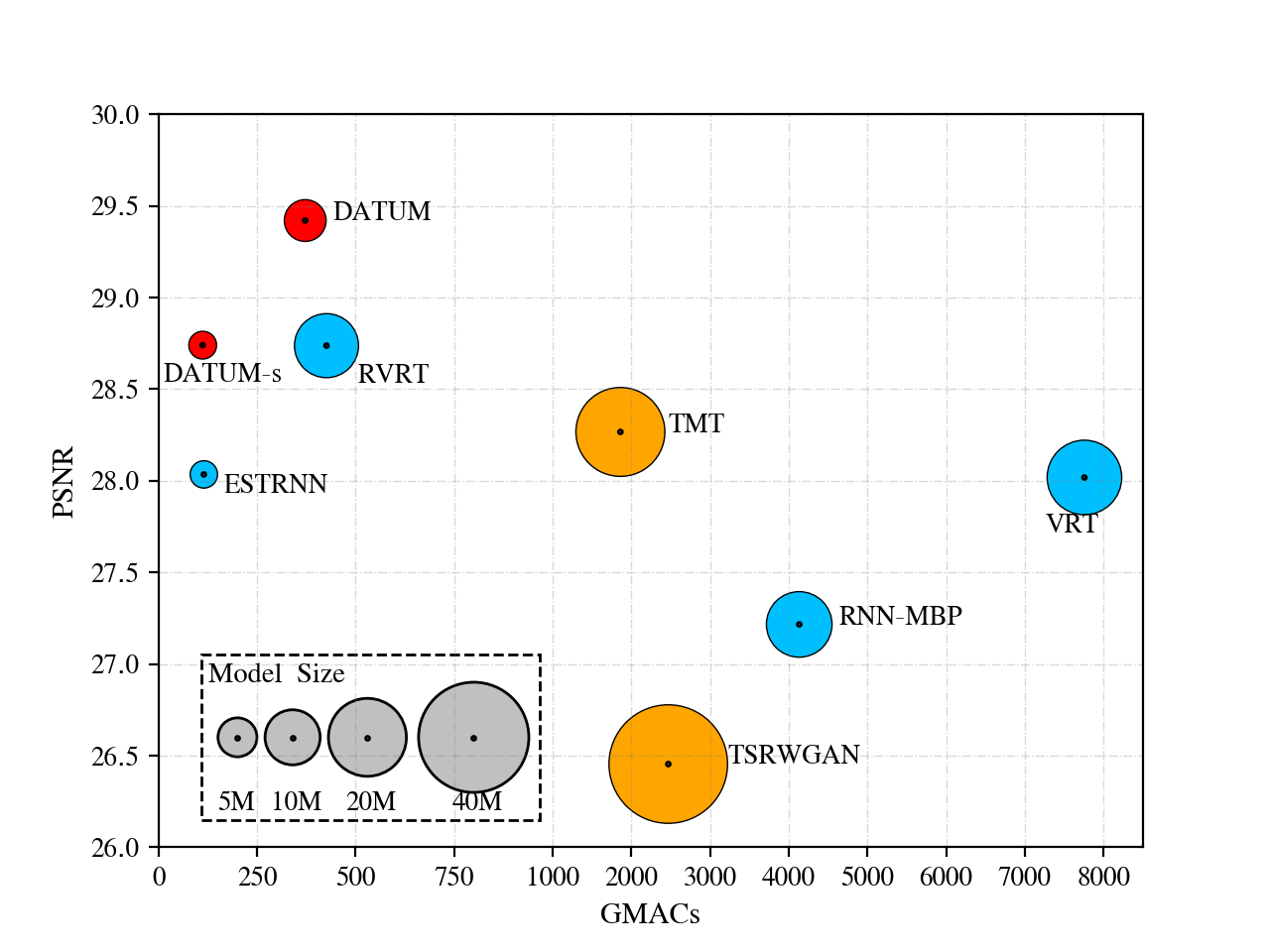}
   \caption{Benchmarking video restoration models for turbulence mitigation on our ATSyn-dynamic dataset. The circles in orange are other video-based TM networks, and the circles in blue are representative video deblurring and general restoration networks. The proposed Deep Atmospheric TUrbulence Mitigation network (DATUM) is state-of-the-art while highly efficient.}
   \label{fig:overall_model}
\end{figure}


With the development of physics-grounded data synthesis methods \cite{Milanfar_2013_a, Lau_2019_a, Chak_2021_a, fraser1999atmospheric, Vorontsov_2001_a, Frakes_2001_a}, data-driven algorithms have been developed in the past two years. Most existing deep learning methods focus on single-frame problems \cite{Lau_2021_a, Rai_2022_a, mei2023ltt, nair2023ddpm, Yasarla_2022_a, Nair_2021_a, Lau_2021_b, Jaiswal_2023_ICCV, Mao_2022_a}. Since the degradation is highly ill-posed, the performance of these algorithms is naturally limited, especially when 
attempting to generalize to real data. On the other hand, multi-frame turbulence mitigation networks \cite{Jin_2021_a, Zhang_2022_a, Anantrasirichai_2022_a} have shown greater potential for generalization across a broader spectrum of real-world test scenarios. However, these networks are adapted from generic video restoration methods and do not reflect the insights developed by traditional methods; few turbulence-specific properties are incorporated as inductive biases into their methods. 


For deep learning methods to work on real-world scenarios, two common factors hinder the application of current turbulence mitigation methods: (1) the complexity of current data-driven methods is usually high, which impedes the practical deployment of these algorithms, and (2) the data synthesis models are suboptimal, either too slow to produce large-scale and diverse datasets or not accurate enough to represent the real-world turbulence profiles, restricting the generalization capability of the model trained on the data.

To overcome these pressing issues, we propose the Deep Atmospheric TUrbulence Mitigation (DATUM) network and the ATSyn dataset. We offer three contributions:
\begin{itemize}
    \item DATUM is the first deep-learning video restoration method customized for turbulence mitigation based on classical insights. By carefully integrating the merits of classical multi-frame TM methods, we propose feature-reference registration, temporal fusion, and the decoupling of pixel rectification and deblurring as effective inductive biases in the multi-frame TM challenge.
    
    \item DATUM is the first recurrent model for turbulence restoration. It is significantly more lightweight and efficient than the prior multi-frame TM methods. On both synthetic and real data, DATUM consistently surpasses the SOTA methods while being $10 \times$ faster.

    \item Through the integration of numerous theoretical and practical improvements in physics modeling over the Zernike-based simulators, we further propose an extensive, real-world inspired dataset ATSyn. Experiments on real-world data show that models trained on ATSyn significantly generalize better than those trained on alternative ones.
\end{itemize}

\section{Related works}
\label{sec:works}
\subsection{Turbulence modeling}

Atmospheric turbulence simulation spans from computational optics to computer vision-oriented approaches. Optical simulations use split-step methods, which numerically propagate waves through phase screens that represent the atmosphere's spatially varying index of refraction \cite{Hardie_2017_a, Roggemann_2012_a, Roggemann_1995_a, Schmidt_2010_a}. Despite the existence of moderately faster optical simulations, including brightness function-based simulations \cite{Vorontsov_2005_a, Lachinova_2007_a, Lachinova_2017_a} or learning-based alternatives \cite{Miller_2019_a, Miller_2021_a}, the relatively slow speed limits their application in deep learning training \cite{Mao_2021_a}. In computer vision simulations, pixels are first displaced according to heuristic correlation functions followed by invariant Gaussian blur \cite{Milanfar_2013_a, Lau_2019_a, Chak_2021_a}, offering speed but arguably lacking physical foundations. 
Recent Zernike-based methods \cite{chimitt2022real, Mao_2021_a, Chimitt_2020_a, chan2023computational} can match the statistics of optics-based simulation, achieving realistic visual quality while maintaining a fast data synthesis speed. It has been applied to turbulence mitigation \cite{Zhang_2022_a, Mao_2022_a, Jaiswal_2023_ICCV, jiang2023nert} to facilitate the generalization capability of those models. 

\subsection{Conventional turbulence mitigation}
Conventional TM algorithms, since \cite{fraser1999atmospheric, Vorontsov_2001_a, Frakes_2001_a}, mostly treat the TM challenge as a many-to-one restoration problem. Considering that turbulence primarily induces random tilt and blur, the common procedure in conventional algorithms is as follows. They first align the input frames to account for pixel displacements, followed by temporal fusion to combine the information from the aligned frames. Subsequently, the residual blur is often considered to be spatially invariant, allowing a blind deconvolution to be applied to produce a visually satisfactory image.

The tilt rectification is typically achieved in a two-step fashion: construct a tilt-free reference frame, then register every frame with respect to the reference. Since the pixel displacement is assumed to be zero-mean over time \cite{Frakes_2001_a, Li_2021_a}, the temporal average can be assumed tilt-free \cite{Shimizu_2008_a, mao_2012_a, Milanfar_2013_a, hardie2021application, Lou_2013_a} and hence be the reference frame. Besides that, low-rank components from all input frames are frequently used \cite{li2007atmospheric, Xie_2016_a, Lau_2019_a} as the reference. The registration step can be done by B-spline or optical flow based warping \cite{Shimizu_2008_a, mao_2012_a,  Milanfar_2013_a, Xie_2016_a, Lou_2013_a} in the spatial domain or phase correction \cite{Anantrasirichai_2013_a, Xue_2016_a, hardie2021application} in the phase domain. Because of the ``lucky effect'' phenomenon \cite{Fried_1978_a} in the short-exposure turbulence, the goal of temporal aggregation is to identify and fuse the randomly emerging sharp regions, a technique known as lucky fusion \cite{Aubailly_2009_a}. \cite{Milanfar_2013_a, Lau_2019_a, Mao_2020_a} design spatial descriptors to select and score lucky regions. \cite{Anantrasirichai_2013_a} identify and fuse sharp components in the wavelet space, and \cite{He_2016_a, Lau_2019_a} apply a similar principle to the sparse components derived through robust PCA. While several methods have been proposed for moving object scenarios \cite{Gepshtein_2004_a, Oreifej_2013_a, Anantrasirichai_2018_a, Nieuwenhuizen_2019_b, Mao_2020_a}, they are restricted by their assumption that the dynamic regions are rigid and can be isolated, leaving the remaining static regions to be restored using the conventional pipelines. 

\begin{figure*}[h]
    \centering
    \includegraphics[width = 0.95\linewidth]{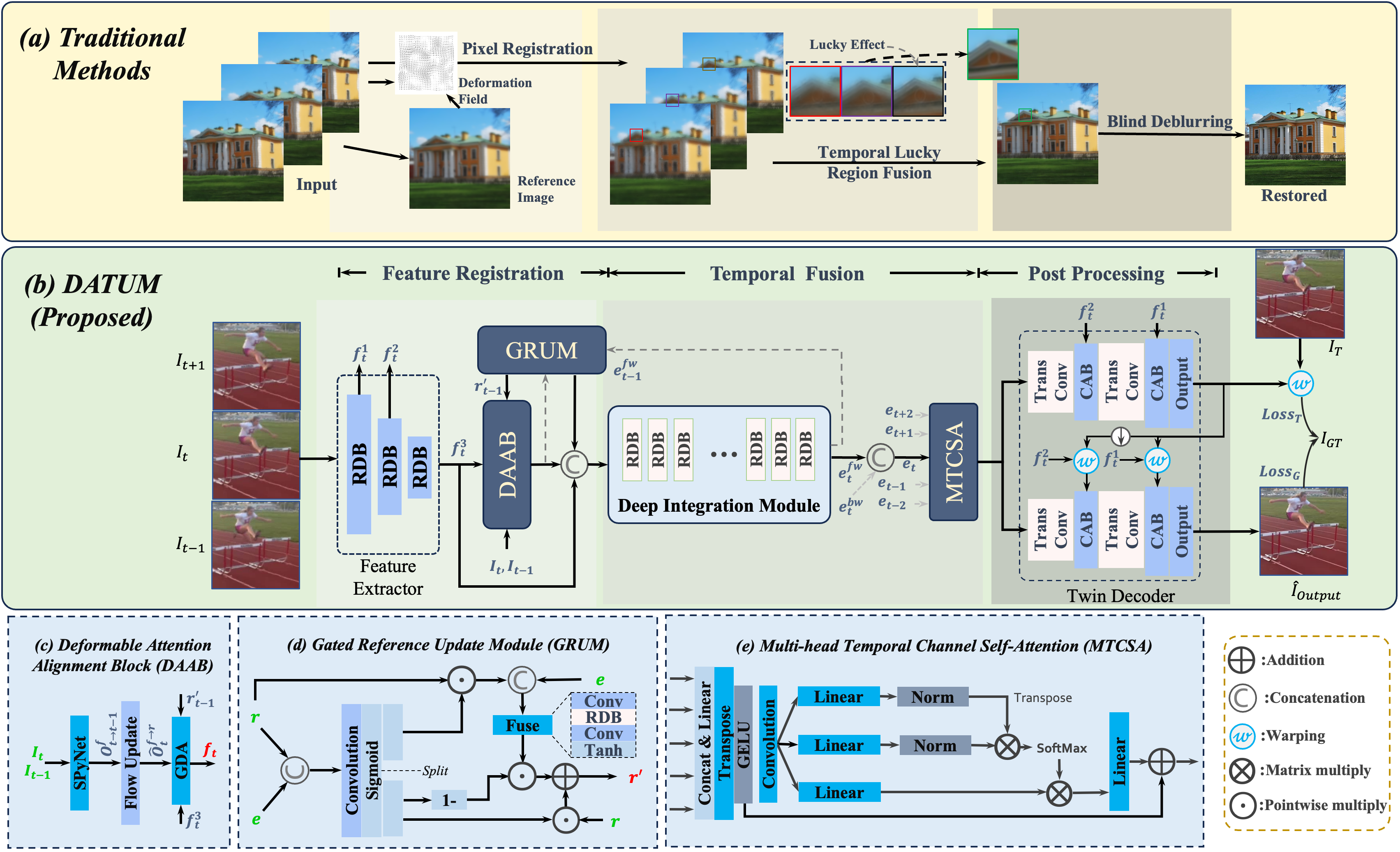}
    \caption{The proposed DATUM network. In this figure, block (a) shows the three common stages proposed by classical TM methods. The corresponding stages in DATUM are shown in block (b), which illustrates the forward time process of the $t\text{-th}$ frame. The dashed line means the information passing from other temporal directions and frames. Block (c), (d), and (e) demonstrate the DAAB, GRUM, and MTCSA modules, respectively, where the \textcolor{green}{input} features are marked by green, and the \textcolor{red}{output} features are marked by red.}
    \label{fig: network}
\end{figure*}

\subsection{Learning-based turbulence mitigation}
 With the rapid advancements in machine learning, numerous recent learning-based methods have demonstrated superior turbulence mitigation results. The majority of them are single-frame TM methods. \cite{Lau_2021_a, Rai_2022_a, mei2023ltt, nair2023ddpm} demonstrate promising performance using generative models with simplified turbulence properties as prior. \cite{Yasarla_2022_a, Nair_2021_a, Lau_2021_b} focus on restoring long-range face images through turbulence. \cite{Jaiswal_2023_ICCV, Mao_2022_a} show physics-grounded synthetic data facilitates certain degrees of generalization capability. These single-frame methods do not account for the temporal dimension and can fall short in multi-frame TM scenarios. In contrast, video-based TM algorithms \cite{Jin_2021_a, Zhang_2022_a} exhibit superior adaptability by leveraging the temporal information, but their designs lack the integration of specific turbulence properties, making their model less efficient. Moreover, \cite{Jin_2021_a} only simulated mild turbulence effect, which restricts the generalization capability of their model. Although \cite{Zhang_2022_a} has achieved better generalization, the point spread function (PSF) implementation is less precise, and the parameter sets are not physics-oriented. Hence, the representative of their turbulence modalities is restricted.

\section{Proposed method}

\subsection{Insights from Classical Methods}
Image degradation by atmospheric turbulence can be roughly described by a compositional operation of the blur $\mathcal{B}$ and the tilt $\mathcal{T}$ via the relationship $\mathbf{I} = [\mathcal{B} \circ \mathcal{T}](\mathbf{J}) + \mathbf{n}$, where $\mathbf{J}$ is the clean image, $\mathbf{I}$ is the distorted image, and $\mathbf{n}$ is the noise term. Traditional algorithms handle turbulence in three steps, as illustrated in Fig. \ref{fig: network}:
\begin{itemize}
\item \textbf{Frame-to-reference registration} \cite{Milanfar_2013_a}, where a reference frame is constructed from the observed images and all images are registered with respect to the reference using optical flow. In strong turbulence or dynamic scenes, constructing a reference is often difficult.
\item \textbf{Lucky image fusion} \cite{joshi_2010_a, Anantrasirichai_2013_a}, where a ``lucky'' image is constructed by collecting the sharpest and most consistent patches from the inputs. However, if turbulence is strong, identifying lucky patches can be difficult.
\item \textbf{Blind deconvolution} \cite{Mao_2020_a}, where a final blind deconvolution algorithm is employed to sharpen the lucky image. The success and failure of this step depend heavily on how spatially uniform the blur in the lucky image is. Oftentimes, since the blur is spatially varying, the performance of blind deconvolution is limited.
\end{itemize}
While each step is important each has its limitations, motivating us to develop end-to-end trained networks to approximate these functions. Empowered by training on our physical-grounded dataset, our network enjoys the inductive biases of those insights while avoiding their limitations.


\subsection{DATUM network}
\label{sec: network}

\subsubsection{Overview}
The block diagram of the DATUM network is depicted in Fig. \ref{fig: network}. We first summarize these three components and describe them in detail in the next subsections.

\textbf{Feature-to-reference registration}. This component is analogous to the classical frame-to-reference registration. For each input frame $I_t$ at time $t$, we first extract three levels of features $f_t^{\{1,2,3\}}$. We propose the Deformable Attention Alignment Block (DAAB) to register the high-level feature $f_t^3$ to a previously hidden reference map $r'_{t-1}$. We also propose the Gated Reference Update Module (GRUM) updates this reference feature recurrently, which is inspired by the gated recurrent unit \cite{CGRU, chung2014empirical} and illustrated in Fig. \ref{fig: network}.

\textbf{Temporal fusion}. This component is analogous to the classical lucky fusion step. The registered feature $f_t$, together with $r'_{t-1}$ and $f_t^3$, are fused by a new Deep Integration Module (DIM). DIM consists of a series of Residual Dense Blocks (RDB) \cite{zhang2018residual} and is used to produce the forward embedding $e_t^{fw}$. Since $e_t^{fw}$ is a deep feature, it is presumed to be free of tilt and is thus utilized for updating the reference feature for the subsequent frame. After the bidirectional recurrent process, we perform a temporal fusion of $e_t^{fw}$ by augmenting it with the backward embedding $e_t^{bw}$ and bidirectional embeddings from neighboring frames. We propose the Multi-head Temporal-Channel Self-Attention (MTCSA) module for this purpose.

\textbf{Post processing}. In the final stage, the temporally fused features are decoded to form the turbulence-free image. This decoding involves a twin of decoders. The first predicts a reverse tilt map that rectifies the shallow features, and the second subsequently reconstructs the clean image.

\subsubsection{Components}

\noindent \textbf{Feature registration via Deformable Attention Alignment Block (DAAB).}
In classical methods, a crucial stage for turbulence mitigation is registering the input frames to the tilt-free reference frame. This reference frame is usually obtained by temporal averaging or using variants of principle component analysis. However, these methods may not be applicable to dynamic videos. Since learning-based video TM is possible \cite{Zhang_2022_a, Jin_2021_a}, the deep feature of a video TM network can be considered tilt-mitigated to work as the reference feature for the next input feature. This section explains our method to use deformable attention to facilitate feature registration in our DATUM network.

The computations in the DDAB are summarized in Algorithm 1, where $\warp(A;B)$ denotes warping $A$ by deformation field $B$, $\phi(A;p)$ denotes sampling $A$ by positions $p$. $W_K$, $W_V$, $W_Q$, and $W$ are linear projections on the channel dimension, and $\sigma$ denotes the SoftMax. The optical flow at line 3 is estimated with the SPyNet \cite{Ranjan_2017_a}, and lines 6-11 are inspired by the guided deformation attention (GDA) \cite{liang2022rvrt}.

\begin{algorithm}
\caption{Deformation Attention Alignment Block}
\begin{algorithmic}[1]
\State \textbf{Input:} Current frame feature $f_t^3$, reference feature $r'_{t-1}$ and alignment flow from last frame $O_{t-1}^{f\rightarrow r}$, two downsampled frames $I_t$ and $I_{t-1}$
\State \textbf{Output:} Updated feature $f_t$ and flow $O_{t}^{f\rightarrow r}$

\Statex \Comment{Estimate rough deformation field $\hat{O}_{t}^{f\rightarrow r}$ that register feature $f_t^3$ to reference $r'_{t-1}$}

\State Estimate the optical flow $O_{t\rightarrow t-1}^{f}$ from $I_t$ and $I_{t-1}$.

\State $\hat{O}_{t}^{f\rightarrow r} \leftarrow O_{t-1}^{f\rightarrow r} + \warp(O_{t\rightarrow t-1}^{f}; O_{t-1}^{f\rightarrow r})$
\State Pre-align $\hat{f}_{t} \leftarrow \warp(\hat{O}_{t}^{f\rightarrow r}, f_{t}^{3})$
\LineComment{Register input feature to reference frame using multi-group multi-head deformation attention}

\ForAll{group $g$}
     \Statex \hskip\algorithmicindent $\triangleright$ Predict offsets $o_{t}^{(g)}$
    \State $\Delta o_{t}^{(g)} \leftarrow \text{RDB}(\text{Concat}(\hat{f}_{t}, r'_{t-1}, \hat{O}_{t}^{f\rightarrow r}))$
    \State $o^{(g)}_{t} \leftarrow \hat{O}_{t}^{f\rightarrow r} + \Delta o_{t}^{(g)}$
    \Statex \hskip\algorithmicindent $\triangleright$ Compute the $g\text{-th}$ aligned feature $\hat{f}_t^{(g)}$:
    \State $K^{(g)} \leftarrow \phi(f_{t}^{3}W_{K}; o^{(g)}_{t})$, $V^{(g)} \leftarrow \phi(f_{t}^{3}W_{V}; o^{(g)}_{t})$
    \State $Q \leftarrow r'_{t-1}W_{Q}$, $\hat{f}^{(g)}_t \leftarrow \sigma(QK^{(g)T} / \sqrt{C}) V^{(g)}$
\EndFor

\State Fuse all groups $f_t \leftarrow \text{Concat}(\{\hat{f}^{(g)}_t\})W$
\State Update final alignment flow $O_{t}^{f\rightarrow r}$ by mean of $\{o^{(g)}_{t}\}$
\State Output $f_t \leftarrow f_t + \text{FeedForward}(f_t)$
\end{algorithmic}
\end{algorithm}

\noindent \textbf{Temporal fusion via Multi-head Temporal Channel Self-Attention (MTCSA).}
After feature registration and deep integration, we propose to augment the embedding with contra-directional information, which is essential to ensure consistent restoration quality across various frames. In addition, like classical methods, a spatially adaptive fusion with adjacent frames is advantageous. We propose the Multi-head Temporal-Channel Self-Attention (MTCSA), as illustrated in Fig. \ref{fig: network}. The MTCSA begins by concatenating channels from multiple frames, followed by a $1 \times 1$ convolution to shrink the channel dimension. Separable convolution is used to construct the spatially varying query, key, and value on the temporal and channel dimensions, and the dynamic fusion is facilitated by self-attention. Finally, a residual connection is used to stabilize training. Considering the quadratic complexity of MTCSA relative to window size, this size is kept moderate. Additionally, we integrate a hard-coded positional embedding wherein features from the focal frame are positioned at the end. This strategy is essential for boundary frames with disproportionate neighboring frames on either side.

\noindent \textbf{Twin decoder and loss function}
 Given the refined feature embedding from the MTCSA, we also developed a twin decoder to progressively remove the tilt and blur, as shown in Fig. \ref{fig: network}. The decoder uses transposed convolution for upsampling and channel attention blocks (CAB) \cite{woo2018cbam} for decoding. Before decoding in higher levels, the deep features are concatenated with the shallow features to facilitate the residual connection like a typical UNet \cite{ronneberger2015u}. Since the deep and shallow features are misaligned by the random tilt $\calL$, we propose to first rectify the shallow features by the estimated inverse tilt field $\hat{\calT}^{-1}$ estimated in the first stage. The tilt-rectification is optimized by reducing the loss:
\begin{equation}
\label{eq:loss_tilt}
    \calL_{\text{tilt}} = \calL_{\text{char}}(\boldsymbol{I}_{\text{GT}}, \warp(\boldsymbol{I}_{\text{tilt}}; \hat{\calT}^{-1}))
\end{equation}
Where $\calL_{\text{char}}$ denotes the Charbonnier loss \cite{charbonnier}, $\boldsymbol{I}_{\text{GT}}$ is the input frame and $\boldsymbol{I}_{\text{tilt}}$ is the tilt-only frame that can be produced without additional cost by our data synthesis method. In the second stage, the rectified shallow features are jointly decoded with the deep features to generate the final reconstruction $\hat{\boldsymbol{I}}$. The overall loss function is computed by:
\begin{equation}
    \calL = \alpha_1 \calL_{\text{tilt}} + \alpha_2 \calL_{\text{char}}(\boldsymbol{I}_{\text{GT}}, \hat{\boldsymbol{I}})
\end{equation}
where weights $\alpha_1$ and $\alpha_2$ are empirically set to 0.2 and 0.8.

\subsection{ATSyn dataset}
\subsubsection{Physics-based data synthesis}
\label{sec: data_synthesis}

\begin{figure}
    \centering
    \includegraphics[width = 0.9\linewidth]{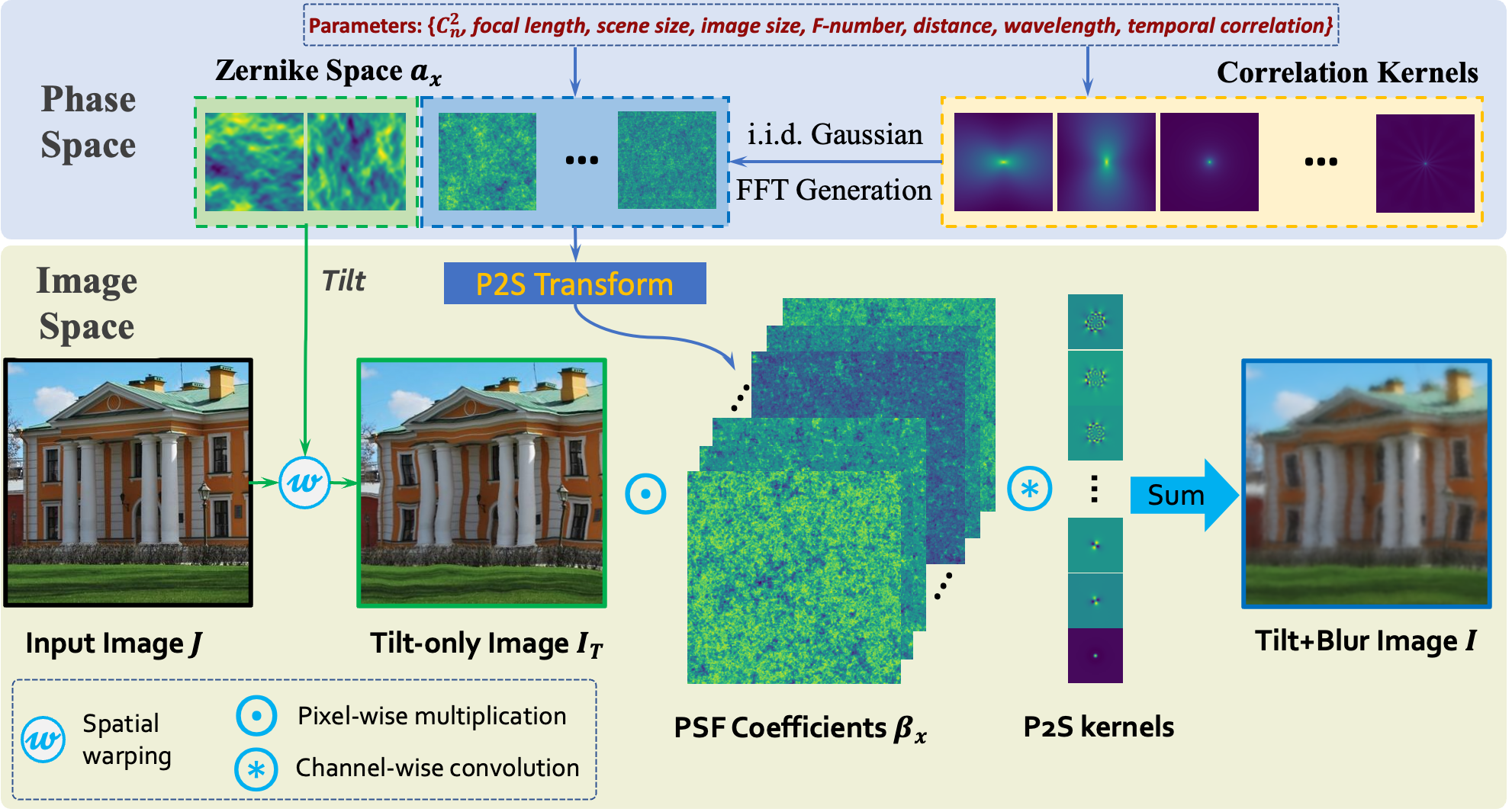}
    \vspace{-0.5em}
    \caption{Scheme of our data synthesis method.}
    \label{fig: synthesis}
\end{figure}

As introduced previously, the ground truth image $\boldsymbol{J}$ is first geometrically distorted and then blurred to produce the degraded image $\boldsymbol{I}$ in our synthesis method. Data synthesis for the turbulence effect essentially requires a physics-grounded representation of $\blur$ and $\tilt$. We adopted the Zernike-based turbulence simulator \cite{Chimitt_2020_a, chimitt2022real} and improved it with non-trivial modifications. Fig. \ref{fig: synthesis} presents the scheme of our implementation. The $\blur$ and $\tilt$ is generated from the phase distortion represented by Zernike polynomials $\{\mZ_i\}$ \cite{Noll_1976_a} as the basis, with corresponding coefficients $\va_{i}$ where $i$ ranging from 1 to 36. Among all 36 coefficients, $i=1$ denotes the current component, $i={2,3}$ controls the $\tilt$ by a constant scale, and the rest high order Zernike coefficients contribute to the blur effect.

 The phase distortion can be assumed as a wide sense stationary (WSS) random field \cite{chimitt2022real}. Hence, it can be sampled with Fast Fourier Transform (FFT) from white Gaussian noise and the autocorrelation map. Transforming the phase distortion to the spatial domain point spread functions (PSF) can be achieved by the Phase-to-Space (P2S) transform, which transforms the sampled Zernike coefficients to spatial coefficients $\boldsymbol{\beta}$, assuming the PSFs can be represented by a low-rank approximation of 100 basis $\boldsymbol{\psi}$ and corresponding $\boldsymbol{\beta}$. The overall degradation in the spatial domain is implemented by 
\begin{equation}
    \boldsymbol{I} = \sum_{k=1}^{100}\boldsymbol{\psi}_k \circledast (\boldsymbol{\beta}_{k} \cdot \warp(\boldsymbol{J}; \tilt)) + \boldsymbol{n},
    \label{eq: p2s_psf_basis}
\end{equation}
where $\circledast$ denotes the depth-wise convolution. Although subtle, this fundamentally generates more reliable degradation than the simulator in \cite{Zhang_2022_a}, as elaborated in \cite{Chimitt_2023_b}. Except for this, our correlation kernels are more precise by incorporating the continuous $C_{n}^{2}$ path technique \cite{Chimitt_2023_a}.

\subsubsection{Guideline of implementation}

With the proposed simulator, we created the ATSyn dataset to match various real-world turbulence conditions and benchmark deep neural networks for turbulence mitigation. This dataset is segmented into two distinct subsets based on scene type: the \emph{ATSyn-dynamic} and \emph{ATSyn-static}. The dynamic sequences contain camera or object motion, whereas the static sequences are each associated with only one underlying clean image. We adopted parameters including focal length, F-number, distance, wavelength, scene size, and sensor resolution to control the simulation. In comparison with the synthetic dataset introduced in \cite{Zhang_2022_a}, which utilized the $D/r_0$ \cite{Fried_1965_a} and empirically chosen blur kernel size, our dataset's parameter space more closely aligns with actual camera settings, making it more representative. 

ATSyn-dynamic contains 4,350 training and 1,097 validation instances synthesized from \cite{Safdarnejad_2015_a, Jin_2021_a}, and ATSyn-static contains 2,000 and 1,000 instances synthesized from the Places dataset \cite{zhou2017places} for training and validation, respectively. Those instances have varying numbers of frames, each with a distinct turbulence parameter set. Besides ground truth and fully degraded videos, ATSyn further provides associated $\boldsymbol{\mathcal{T}}$-only videos to facilitate the training of $\calL_{\text{tilt}}$ in Eq. \ref{eq:loss_tilt}. We categorize the turbulence parameters by three levels: \emph{weak}, \emph{medium}, and \emph{strong}. The range of turbulence parameters is determined by matching with a large-scale, long-range video dataset \cite{cornett2023expanding} and other real-world videos, with more details in the section \ref{dataset}.

\section{Experiments}

\begin{table*}
\centering
\resizebox{0.98\textwidth}{!}{
\begin{tabular}{c|cccccccc}
\hline
Methods & TurbNet \cite{Mao_2022_a} & TSRWGAN \cite{Jin_2021_a} & VRT \cite{liang2022vrt}  & TMT \cite{Zhang_2022_a} & RNN-MBP \cite{zhu2022deep} & ESTRNN \cite{zhong2022real} & RVRT \cite{liang2022rvrt} & DATUM [ours]  \\
\hline
PSNR &  24.2229 & 26.3262 & 27.6114 & 27.7419 & 27.7152 & 27.3469 &  27.8512 & \textbf{28.5875} \\
$\text{SSIM}_{\text{CW}}$ &  0.8230 & 0.8596 & 0.8691 & 0.8741 & 0.8730 & 0.8617 & 0.8788 & \textbf{0.8803}  \\
\hline
\end{tabular}
}
\vspace{-0.5em}
\caption{Preliminary study: evaluate on TMT's synthetic dynamic scene data \cite{Zhang_2022_a}. $\text{SSIM}_{\text{CW}}$ denotes Complex Wavelet SSIM.}
\label{table:pre}
\end{table*}

\begin{table*}
\centering
\setlength{\aboverulesep}{0pt}
\setlength{\belowrulesep}{0pt}
\resizebox{0.98\textwidth}{!}{
\begin{tabular}{l|cccccccc|cc}
\toprule[0.8pt]
    Turbulence Level & \multicolumn{2}{c}{Weak} & \multicolumn{2}{c}{Medium} & \multicolumn{2}{c}{Strong}  & \multicolumn{2}{c|}{Overall} & \multicolumn{2}{c}{Cost}  \\
    \hline
Methods & PSNR & $\text{SSIM}_{\text{CW}}$ & PSNR & $\text{SSIM}_{\text{CW}}$ & PSNR & $\text{SSIM}_{\text{CW}}$ & PSNR & $\text{SSIM}_{\text{CW}}$ & Size  & FPS \\
\hline
TSRWGAN \cite{Jin_2021_a} & 27.0844 & 0.8575 & 26.7046 & 0.8514 & 25.4230 & 0.8372 &  26.4541 & 0.8493 & 46.28 & 0.87 \\
TMT \cite{Zhang_2022_a} & 29.1183 & 0.8836  & 28.5050 & 0.8791  & 26.9744 & 0.8552  & 28.2665 & 0.8734  & 26.04 &  0.80\\
VRT \cite{liang2022vrt} & 28.8453 &  0.8797 & 28.2628 & 0.8769  & 26.7492 & 0.8506  & 28.0179 &  0.8699 & 18.32 & 0.17\\
RNN-MBP \cite{zhu2022deep} & 27.9243 &  0.8699  & 27.4742 & 0.8642 & 26.0812 & 0.8495 & 27.2161 & 0.8618 & 14.16 &  1.14 \\
ESTRNN \cite{zhong2022real} & 28.9805 & 0.8750  & 28.3338 & 0.8697   & 26.8897 & 0.8463  & 28.1347 & 0.8645 & 2.468 & 27.65\\
RVRT \cite{liang2022rvrt} & 29.6080 &  0.8845 & 28.9605 & 0.8806 & 27.5344 &  0.8595 & 28.7672 & 0.8756 & 13.50 & 2.43 \\
\rowcolor{Gray}DATUM-s [ours] & 29.5958 & 0.8809 & 28.9869 & 0.8762 & 27.5456 & 0.8550 & 28.7743 & 0.8714 & 2.538 & 22.48 \\
\rowcolor{Gray}DATUM [ours] & \textbf{30.2058} & \textbf{0.8857} & \textbf{29.6203} & \textbf{0.8829} & \textbf{28.2550} & \textbf{0.8640} & \textbf{29.4222} & \textbf{0.8781}  & 5.754 & 9.17 \\
\bottomrule[0.8pt]
\end{tabular}
}
\caption{Performance comparison on the ATSyn-dynamic set, we list the image quality scores on different turbulence levels and frame-wise resource consumption (measured with $960\! \times \! 540$ frame sequences on RTX 2080 Ti).}
\label{table:dynamic}
\end{table*}

\subsection{Training setting}
This section describes how we trained our DATUM and other models. Except for turbulence mitigation networks \cite{Jin_2021_a, Zhang_2022_a, Mao_2022_a}, we also benchmarked several representative video restoration \cite{liang2022vrt, liang2022rvrt} and deblurring networks \cite{zhong2022real,zhu2022deep} for a more thorough comparison.

To train the proposed model, we used the Adam optimizer \cite{kingma2014adam} with the Cosine Annealing learning rate schedule \cite{loshchilov2016sgdr}. The initial learning rate is $2\times 10^{-4}$, and batch size is 8. All dynamic scene TM networks in this experiment are trained end-to-end from scratch for 800K iterations. To get their static-scene variant, we fine-tuned them on the static-scene modality with half the initial learning rate and 400K iterations. We clip the gradient if the L2 norm exceeds 20 to prevent gradient explosion during inference.

We trained the ESTRNN \cite{zhong2022real}, RNN-MBP \cite{zhu2022deep}, and RVRT \cite{zhong2022real} with the same configuration as DATUM. The number of input frames of DATUM and ESTRNN during training is set to 30 for ATSyn-dynamic and 36 for ATSyn-static. Since RNN-MBP and RVRT require much more resources to train, the number of input frames is set to 16. Because TSRWGAN \cite{Jin_2021_a}, TMT \cite{Zhang_2022_a}, and TurbNet \cite{Mao_2022_a} are all designed for turbulence mitigation, we trained them following the original paper and public code. 

\subsection{Comparison on dynamic scene modality}
We first trained and evaluated all networks for comparison on a previous Zernike-based synthetic dataset \cite{Zhang_2022_a} for preliminary study. We choose PSNR and Complex Wavelet Structure Similarity \cite{sampat2009complex} (CW-SSIM) as the criterion in this paper, and the reason for selecting CW-SSIM rather than SSIM is provided in the section \ref{metrics}. The result in Table \ref{table:pre} shows our DATUM outperforms the previous state-of-the-art TMT \cite{Zhang_2022_a} with $5\times$ fewer parameters and over $10 \times$ faster inference speed. We also benchmark a representative single-frame TM network \cite{Mao_2022_a} to demonstrate the superiority of multi-frame TM methods. 

Next, we present extensive results from the ATSyn-dynamic dataset in Table \ref{table:dynamic}. Our model outperforms all other networks by a significant margin, while it is the second smallest network among all models and the most efficient network among all existing turbulence mitigation networks. To further substantiate the efficacy of DATUM’s design, we introduced a scaled-down variant, DATUM-s.  Although DATUM-s retains the fundamental architecture of DATUM, it operates with only half the number of channels. This reduction assesses the model’s performance under constrained computational resources, offering insights into its scalability and efficiency.


\subsection{Comparison on static scene modality}

When training on the ATSyn-static, the loss is computed between the single ground truth and all output frames. For testing, we instead calculate the average score of the central four frames in the entire output sequence (for single-directional models, we use the last 4). We evaluated the performance on the ATSyn-static and the turbulence text dataset \cite{UG2}, and the result is shown in Table \ref{table:static}. The turbulence text dataset contains 100 sequences of text images, each a static scene of degraded text pattern captured at 300 meters or farther. Real-world turbulence videos do not have ground truth, while \cite{UG2} uses the accuracy score of pre-trained text recognition models CRNN \cite{shi2016end}, DAN \cite{wang2020decoupled}, and ASTER \cite{shi2018aster} as metrics, where a better turbulence mitigation offers better recognition performances. Our model is trained on a wide range of turbulence conditions and generic data, without specific augmentation tricks, yet performs on par with the best systems in the $\text{UG2+}$ turbulence challenge \cite{UG2}. Our model outperforms other networks trained on the ATSyn-static dataset by an even larger margin.

\begin{table}
\centering
\small
\setlength{\aboverulesep}{0pt}
\setlength{\belowrulesep}{0pt}
\resizebox{0.47\textwidth}{!}{
\begin{tabular}{lccc}
\toprule[1pt]
Benchmark & \multicolumn{2}{c}{ATSyn-static} & Turb-Text ($\%$) \\
 \cmidrule(r){1-1} \cmidrule(r){2-3} \cmidrule(r){4-4} 
 Methods & PSNR & $\text{SSIM}_{\text{CW}}$ & CRNN/DAN/ASTER \\
\midrule
TSRWGAN \cite{Jin_2021_a} & 23.16 & 0.8407 & 60.30 / 73.90 / 74.40 \\ 
TMT \cite{Zhang_2022_a} & 24.51 & 0.8716 & 80.90 / 87.25 / 88.55 \\
VRT \cite{liang2022vrt} & 24.27 & 0.8641 & 76.30 / 84.45 / 83.60 \\
RNN-MBP \cite{zhu2022deep} & 24.64 & 0.8775 & 51.35 / 65.00 / 64.30\\
ESTRNN \cite{zhong2022real} & 26.23 & 0.9017 & 87.10 / 97.80 / 96.95 \\
RVRT \cite{liang2022rvrt} & 25.71 & 0.8876 & 86.40 / 89.00 / 89.20 \\
\rowcolor{Gray} DATUM [ours]  & \textbf{26.76} & \textbf{0.9102} & \textbf{93.55 / 97.95 / 97.25} \\
\bottomrule[1pt]
\end{tabular}
}
\caption{Static scene modality. CRNN/DAN/ASTER are the text recognition rates of these three models from the restored images.}
\label{table:static}
\vspace{-0.5em}
\end{table}

\subsection{Ablation study}
Our ablation study examines key elements that introduce effective inductive biases of our model, including the use of additional frames, recurrent reference updating, feature-reference registration, and multi-frame embedding fusion.

\noindent \textbf{Influence of the number of input frames.}
The number of input frames for both training and inference matters for recurrent-based networks, especially in turbulence mitigation. Since turbulence degradation is caused by zero-mean stochastic phase distortion, the more frames the network can perceive, the better the non-distortion state it can evaluate. This is particularly valid for static scene sequences, where the pixel-level turbulence statistics are much easier to track and analyze through time.

\begin{figure}
    \centering
    \footnotesize
    \begin{tabular}{cc}
        \includegraphics[width = 0.45\linewidth]{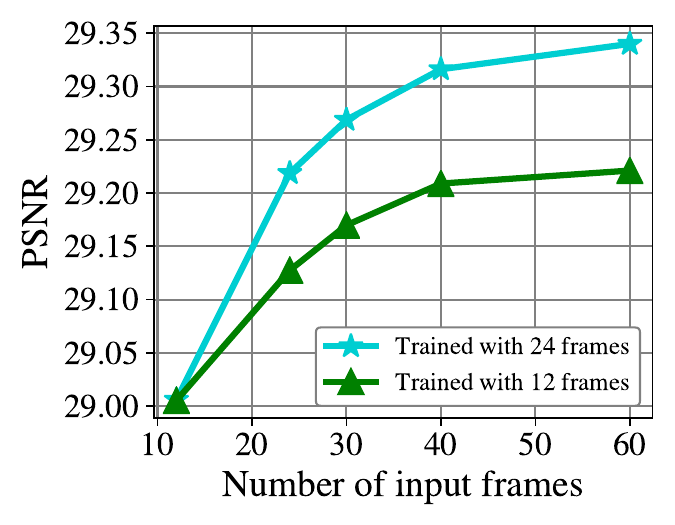} &
        \includegraphics[width = 0.45\linewidth]{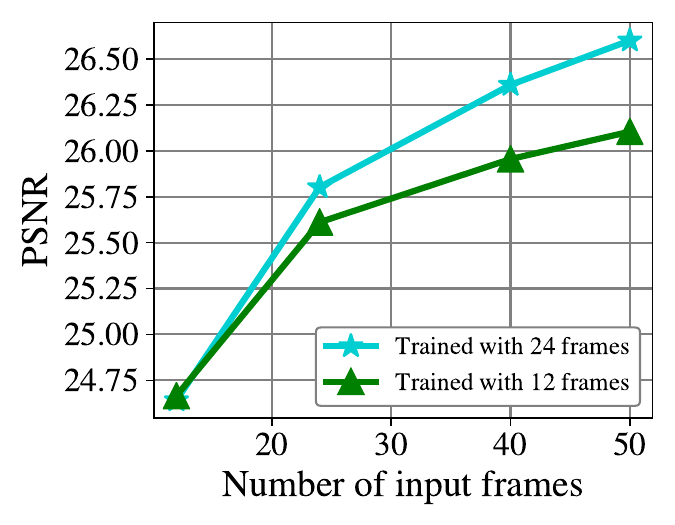} \\
        (a) On ATSyn-dynamic &  (b) On ATSyn-static
    \end{tabular}
    \vspace{-0.5em}
    \caption{Influence of the number of input frames.}
    \label{fig: temporal_len}
\end{figure}

We trained two models with 12-frame and 24-frame inputs and presented their respective performance during inference in Fig. \ref{fig: temporal_len}. This figure shows in the temporal range of our experimental setting, a positive correlation between the performance and the number of input frames always exists, especially on the static scene modality where an over 1 dB boost can be obtained with more frames. This phenomenon suggests one of the success factors for turbulence mitigation is the capability of fusing more frames, similar to the video super-resolution problem \cite{chan2022investigating}. 

\noindent \textbf{Influence of DAAB, MTCSA, GRUM, and twin decoder.}
The design of DAAB and MTCSA are inspired by pixel registration and lucky fusion in the conventional TM methods. Although our spatial registration and temporal fusion are implemented at the feature level, they are still effective in turbulence mitigation, as shown in Table \ref{table:ablation}.

While the MTCSA fuses embeddings from multiple frames in a sliding window manner, determining the optimal window size is crucial. If the window size is too small, the temporal fusion only relies on the implicit temporal propagation by the recurrent unit, limiting the performance; if the window size is too large, because of the quadratic complexity along the temporal dimension, the MTCSA becomes very resource-demanded, and the network becomes less flexible to deal with a small number of input frames. We investigated the temporal window size of the MTCSA module, as shown in Fig. \ref{table:ablation}, where we found that five frames meet the trade-off between performance and efficiency. 

The GRUM utilizes a gating mechanism in the recurrent network to facilitate more extended temporal dependency \cite{CGRU, chung2014empirical}. It fuses the reference feature with deeper embeddings in a more adaptive manner, which also turns out to be effective. Finally, in the post-processing stage, we compared the two-stage twin decoder with the one-stage plain decoder. We found that by incorporating additional supervision and rectifying shallow features in the decoding stage, better performance can be obtained.

\begin{table}
\centering
\resizebox{0.47\textwidth}{!}{
\setlength{\aboverulesep}{0pt}
\setlength{\belowrulesep}{0pt}
\begin{tabular}{lccc}
\toprule[0.8pt]
Components & PSNR / SSIM & Size & GMACs \\
\midrule
\hspace{0.5em} Base (MTCSA-1f)  &  28.62 / 0.8465 & 3.912 & 261.5 \\
\hspace{0.5em} Base (MTCSA-3f) & 28.79 / 0.8497 & 4.131 & 272.7 \\ 
$\ast$ Base (MTCSA-5f) & 28.87 / 0.8522 & 4.768  & 304.2 \\ 
\hspace{0.5em} Base (MTCSA-7f) & 28.92 / 0.8532 & 5.808 & 358.1 \\ 

+ GRUM  &  29.06 / 0.8576 & 4.894 & 317.7  \\
+ DAAB & 29.33 / 0.8638 & 5.241 & 351.8 \\ 
+ Twin Decoder & 29.42 / 0.8647 & 5.754 & 372.7 \\ 
\bottomrule[0.8pt]
\end{tabular}
}
\caption{Ablation study. We conducted experiments on the ATSyn-dynamic set by adding each proposed component progressively and observed a constant performance improvement.}
\label{table:ablation}
\end{table}

\begin{figure}
    \centering
    \includegraphics[width = 0.9\linewidth]{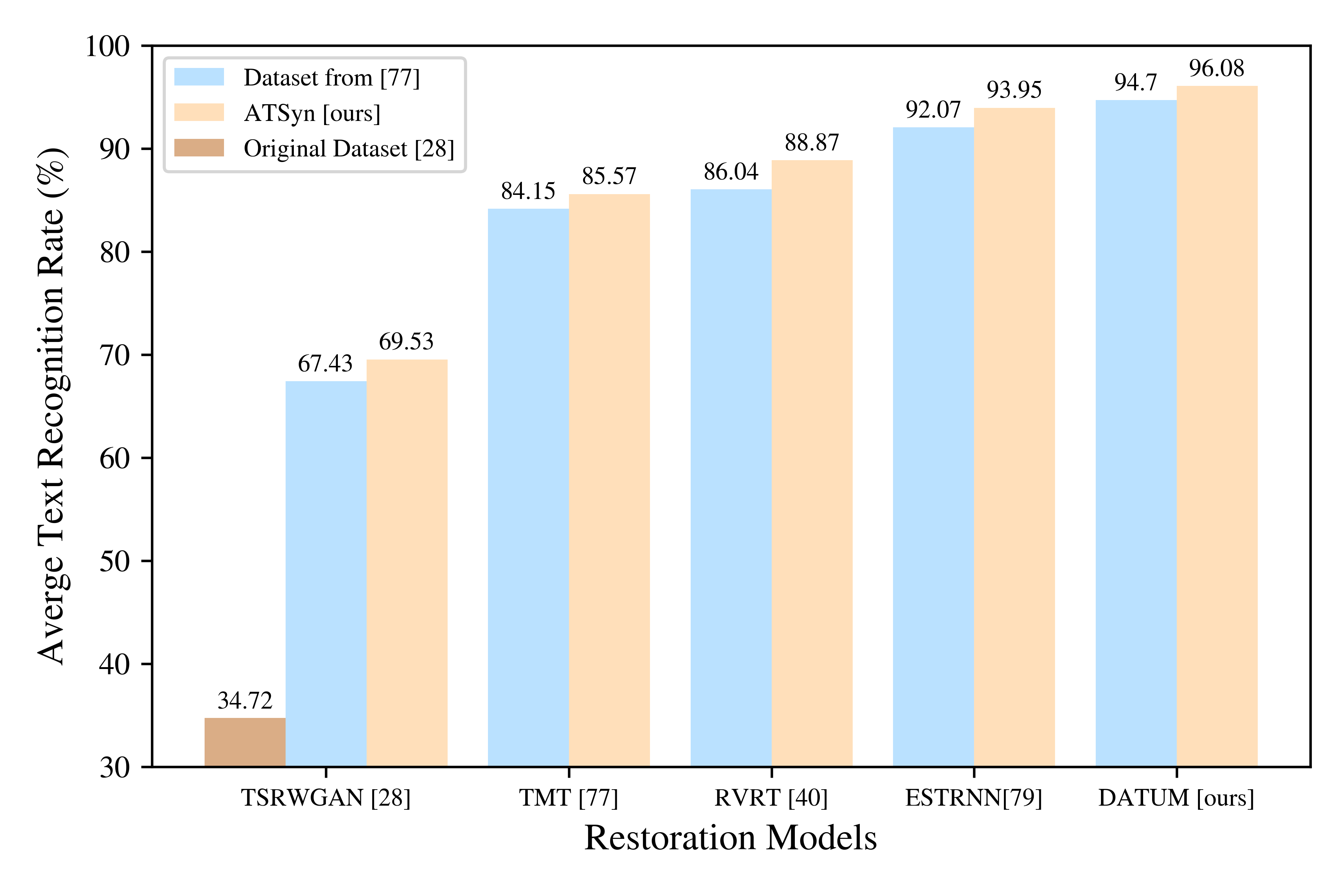}
    \vspace{-0.5em}
    \caption{Comparison on the real-world turbulence-text dataset. The metric is the average text recognition accuracy of CRNN, DAN, and ASTER tested on the restored images.}
    \label{fig:quant_text}
\end{figure}

\subsection{Comparison on real-world data}

\begin{figure*}[ht]
    \captionsetup[subfloat]{font=scriptsize}
    \centering
  \subfloat[Input frame]{%
      \includegraphics[width=0.14\linewidth]{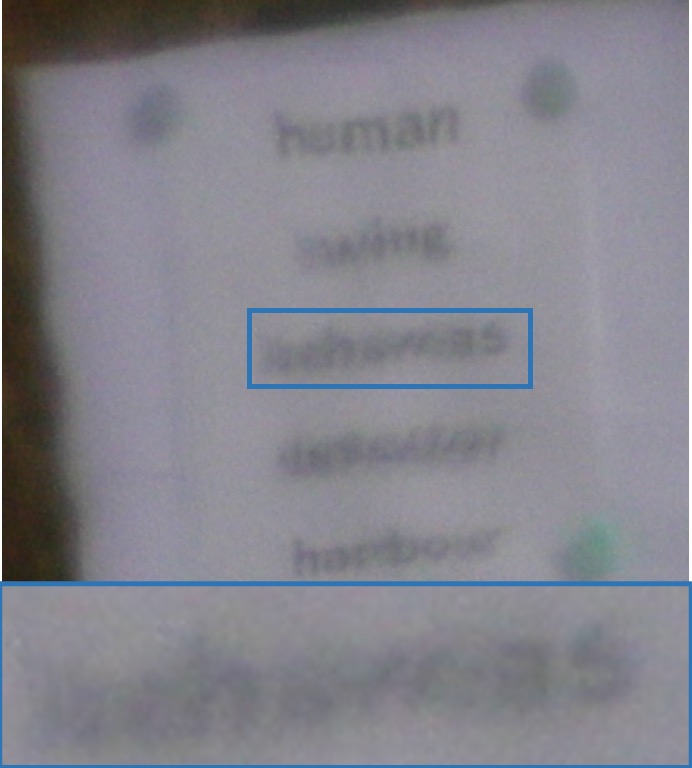}}
    \hfill
  \subfloat[TSRWGAN \cite{Jin_2021_a}]{%
    \includegraphics[width=0.14\linewidth]{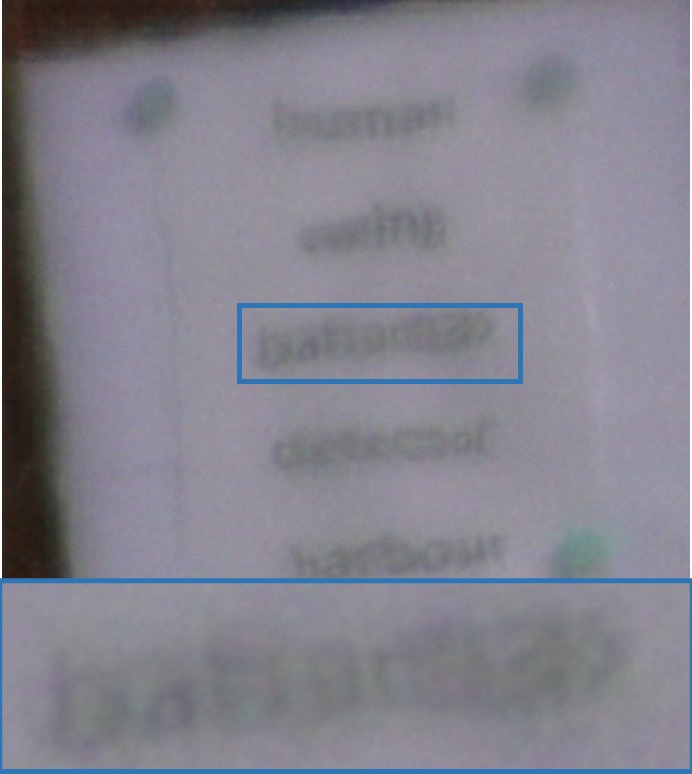}}
   \hfill
  \subfloat[TMT \cite{Zhang_2022_a}]{%
        \includegraphics[width=0.14\linewidth]{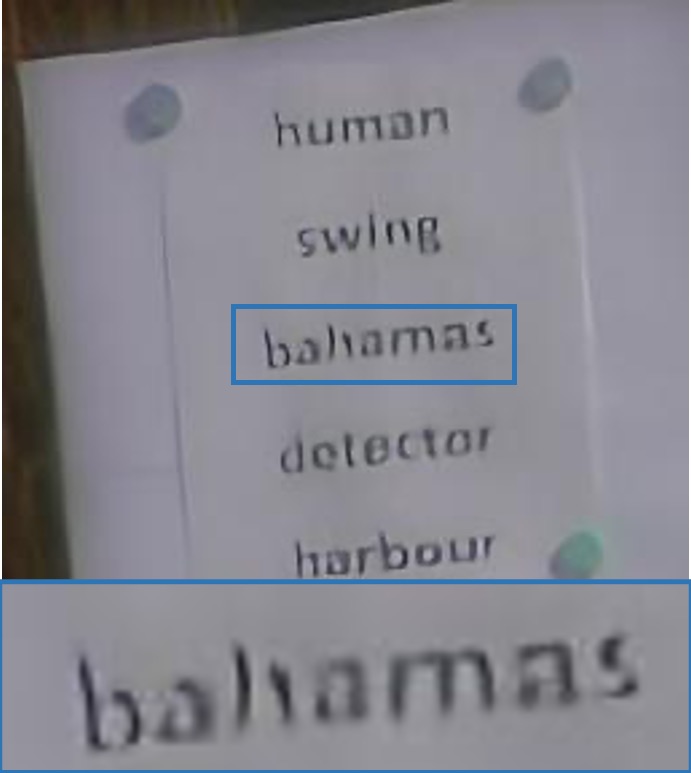}}
    \hfill
  \subfloat[NDIR \cite{Li_2021_a}]{%
      \includegraphics[width=0.14\linewidth]{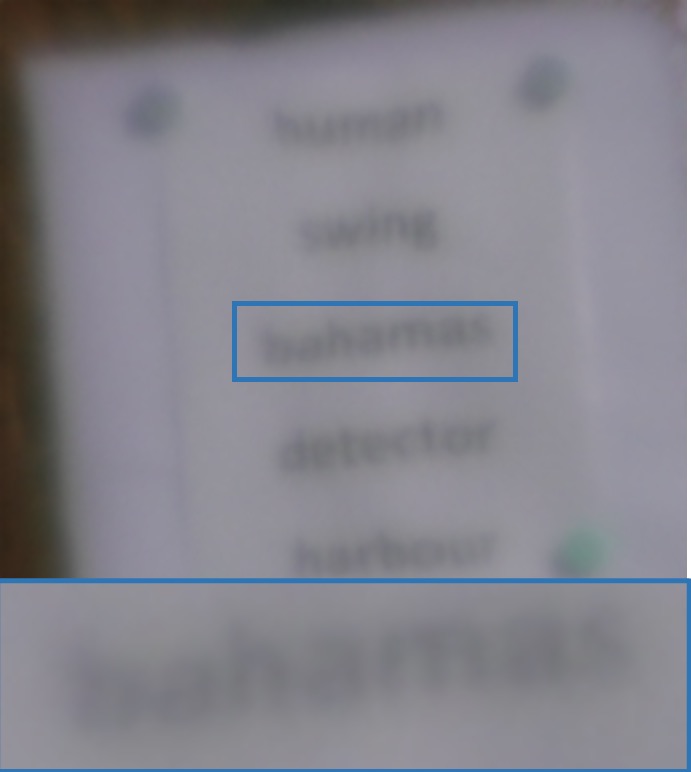}}
    \hfill
  \subfloat[TurbNet \cite{Mao_2022_a}]{%
        \includegraphics[width=0.14\linewidth]{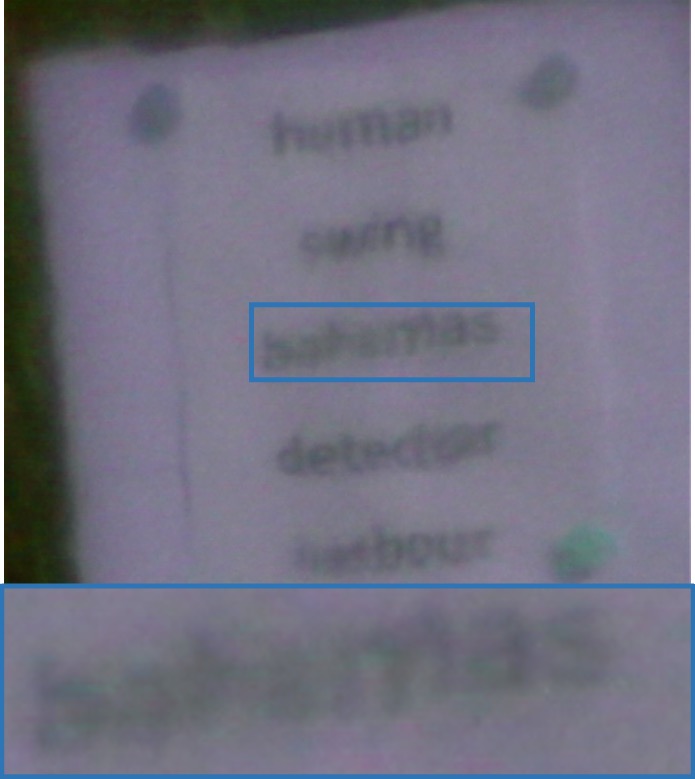}}
    \hfill
  \subfloat[AT-DDPM \cite{nair2023ddpm}]{%
      \includegraphics[width=0.14\linewidth]{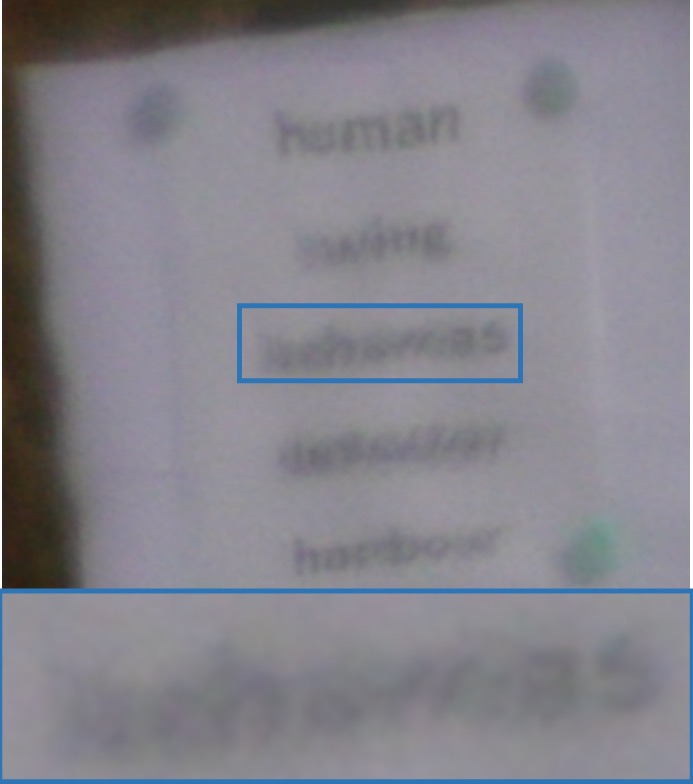}}
    \hfill
  \subfloat[PiRN \cite{Jaiswal_2023_ICCV}]{%
      \includegraphics[width=0.14\linewidth]{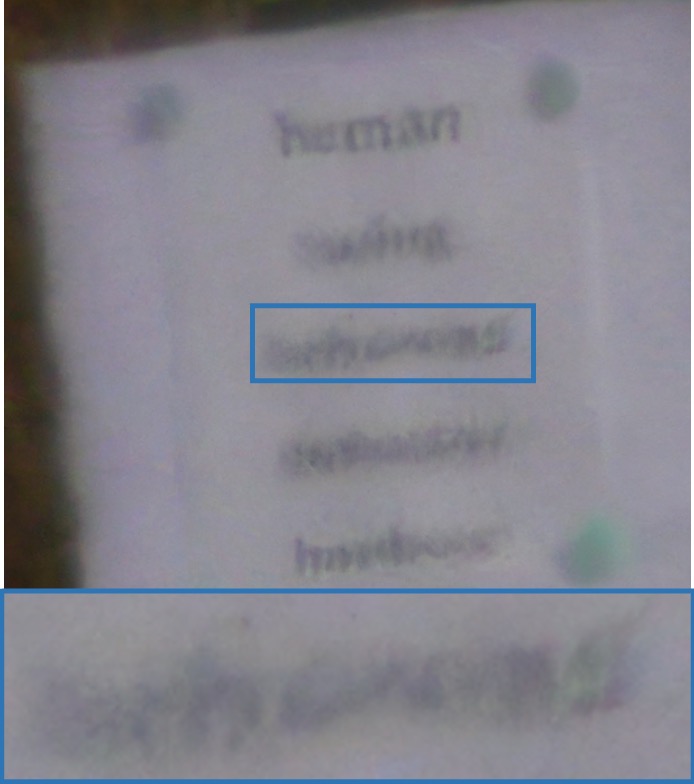}}
    \hfill
    
  \subfloat[\textbf{DATUM [Ours]}]{%
      \includegraphics[width=0.14\linewidth]{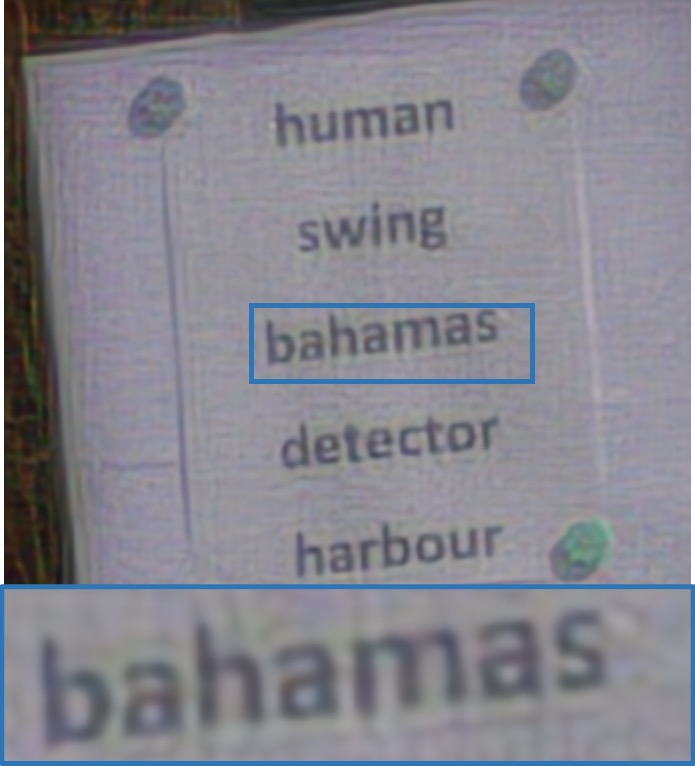}}
    \hfill
  \subfloat[TSRWGAN* \cite{Jin_2021_a}]{%
    \includegraphics[width=0.14\linewidth]{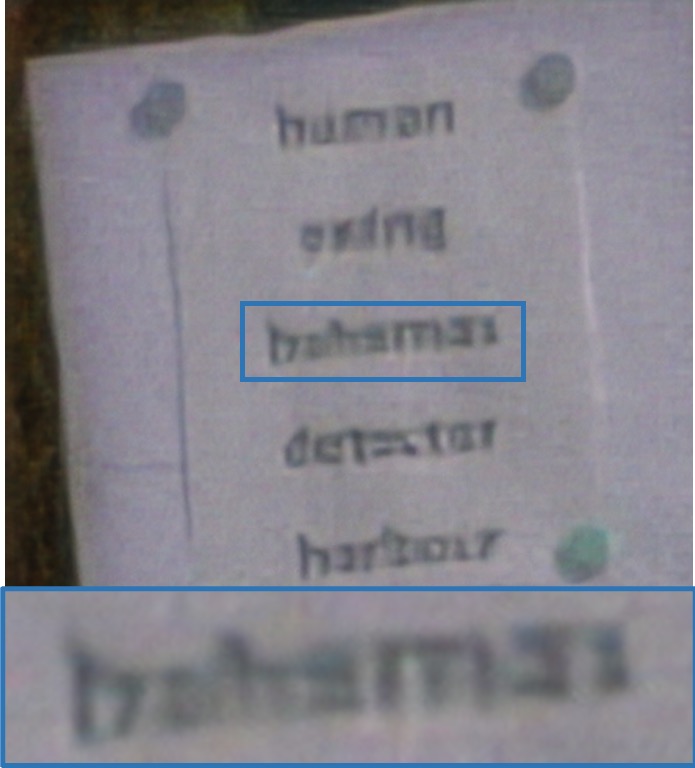}}
   \hfill
  \subfloat[TMT* \cite{Zhang_2022_a}]{%
        \includegraphics[width=0.14\linewidth]{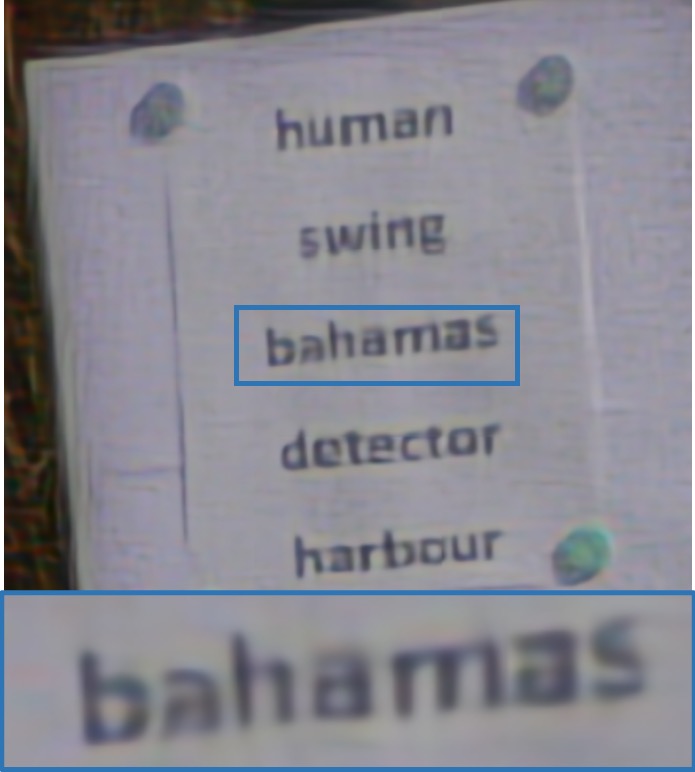}}
    \hfill
  \subfloat[RNN-MBP* \cite{Mao_2022_a}]{%
        \includegraphics[width=0.14\linewidth]{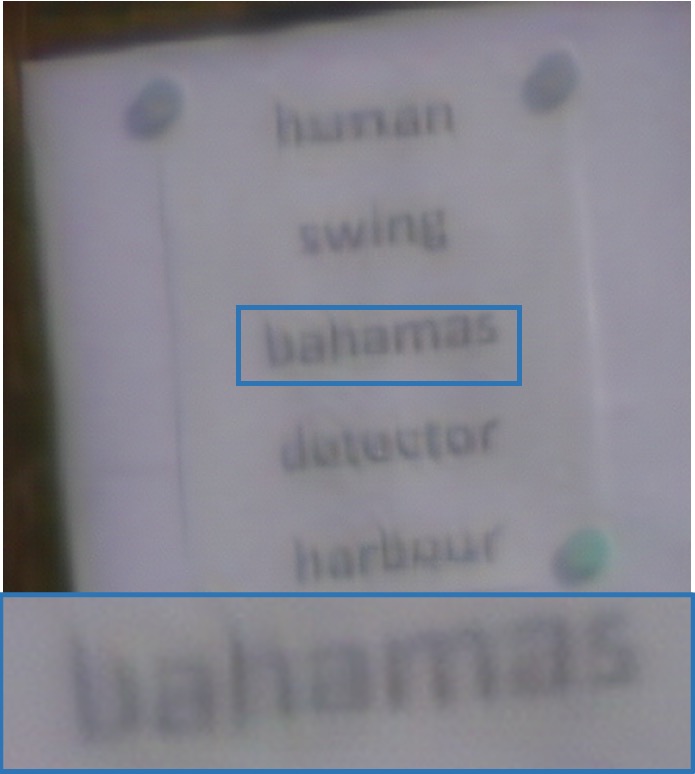}}
    \hfill
  \subfloat[RVRT* \cite{liang2022rvrt}]{%
      \includegraphics[width=0.14\linewidth]{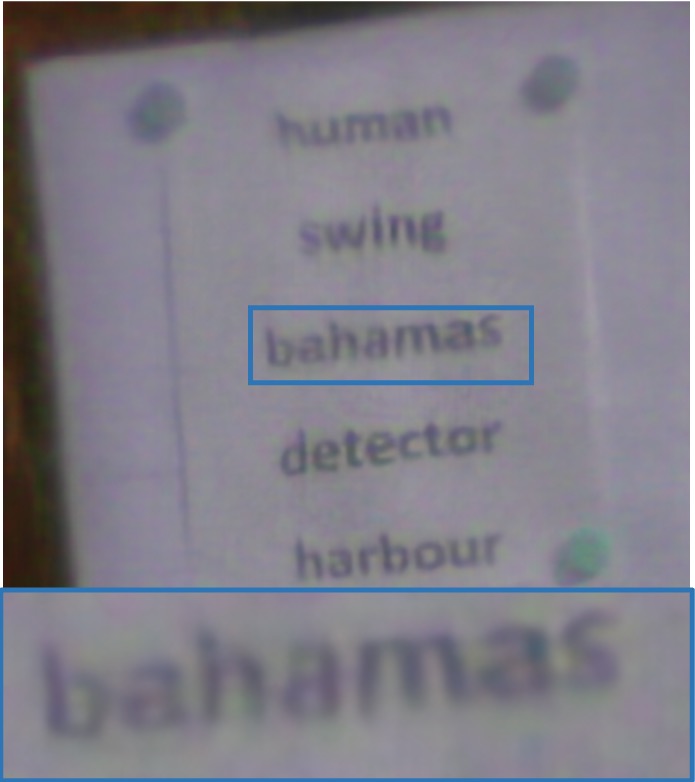}}
    \hfill
  \subfloat[VRT* \cite{Jin_2021_a}]{%
      \includegraphics[width=0.14\linewidth]{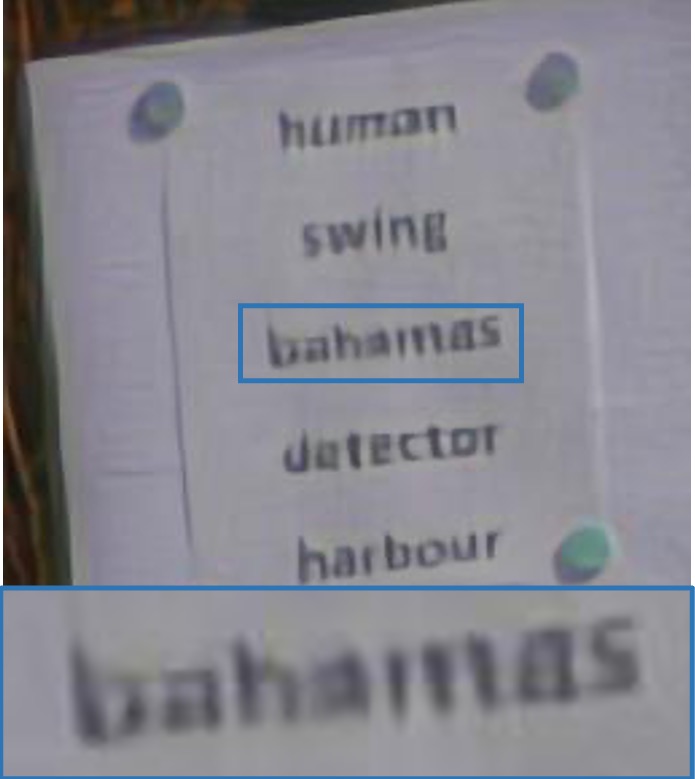}}
    \hfill
  \subfloat[ESTRNN* \cite{zhong2022real}]{%
      \includegraphics[width=0.14\linewidth]{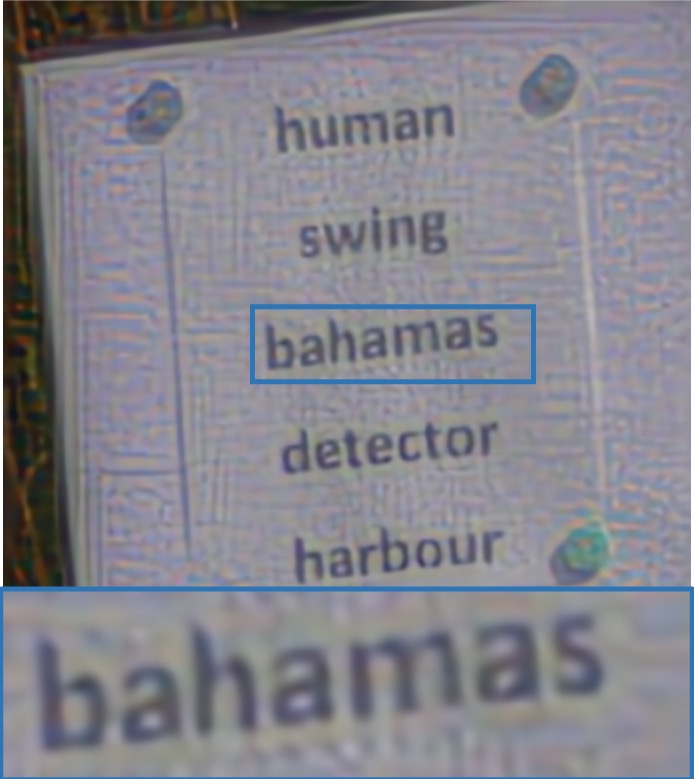}}
    \hfill

  \caption{Qualitative comparison on the turbulence-text dataset \cite{UG2}. The input frame (a) is the 49th frame of the 94th sequence in \cite{UG2}. Figures on the top row are restoration results of corresponding TM methods using their original model and checkpoints. Figures on the bottom row are TM or general restoration models (marked by *) trained on our \emph{ATSyn-static} dataset.}
  \label{fig:text} 
\end{figure*}


\begin{table}
\centering
\setlength{\aboverulesep}{0pt}
\setlength{\belowrulesep}{0pt}
\resizebox{0.46\textwidth}{!}{
\begin{tabular}{c|ccc}
\hline
 Face Retrieval  & Degraded &  Simulator in \cite{Zhang_2022_a} &  Our simulator \\ \hline
 Rank 5  & 37.75\%       &  38.83\%  &  \textbf{39.18\%}        \\
Rank 10 & 40.59\%        & 41.83\% & \textbf{42.18\%}        \\
Rank 20 & 45.29\%      &  46.40\%  & \textbf{46.70\%} \\
\hline
\end{tabular}}
\caption{Face recognition results on a subset of the BRIAR dataset.}
\vspace{-0.5em}
\label{tab:face}
\end{table}

In this section, we demonstrate our data's generalization capability qualitatively and quantitatively on real-world data. 

\begin{figure*}[ht]
    \captionsetup[subfloat]{font=scriptsize}
    \centering
  \subfloat[Input frame]{%
      \includegraphics[width=0.14\linewidth]{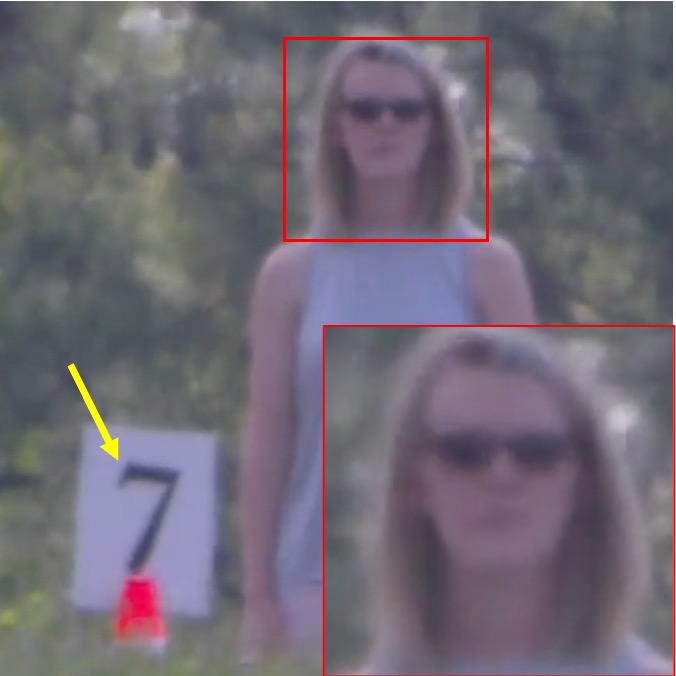}}
    \hfill
  \subfloat[TSRWGAN \cite{Jin_2021_a}]{%
    \includegraphics[width=0.14\linewidth]{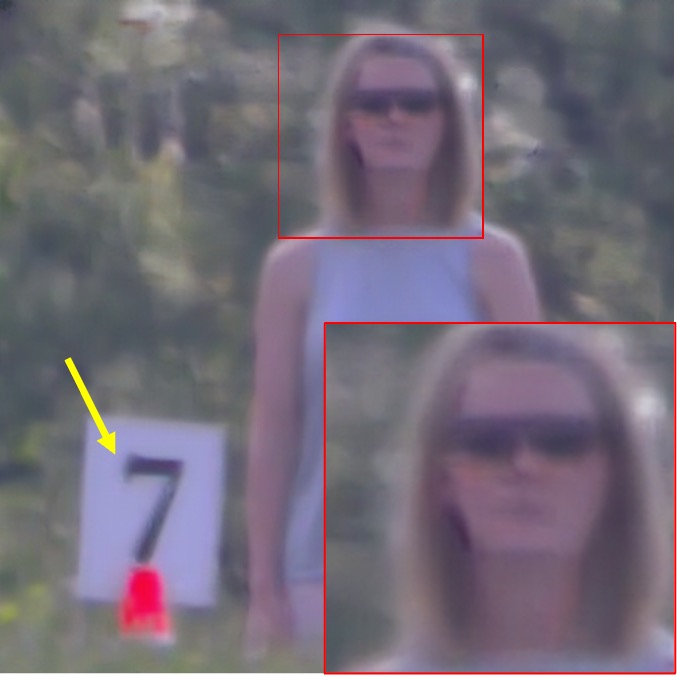}}
   \hfill
  \subfloat[TMT \cite{Zhang_2022_a}]{%
        \includegraphics[width=0.14\linewidth]{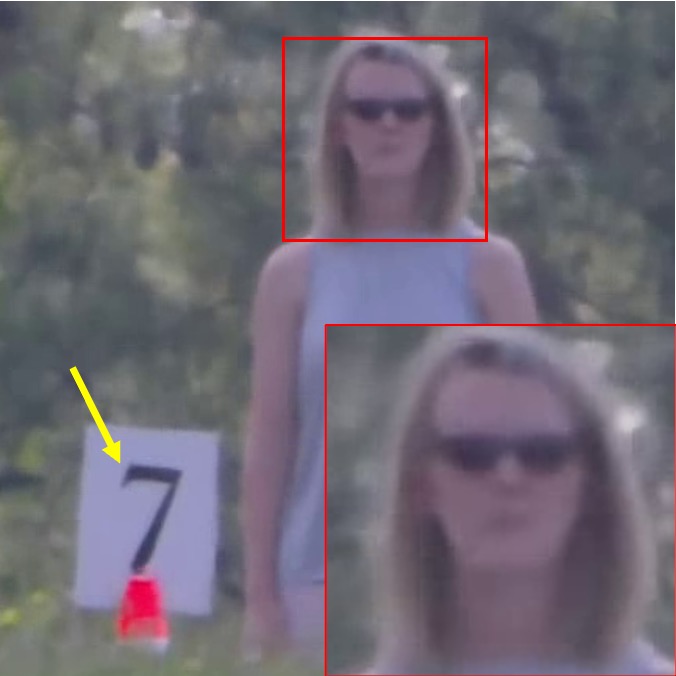}}
    \hfill
  \subfloat[TurbNet \cite{Mao_2022_a}]{%
        \includegraphics[width=0.14\linewidth]{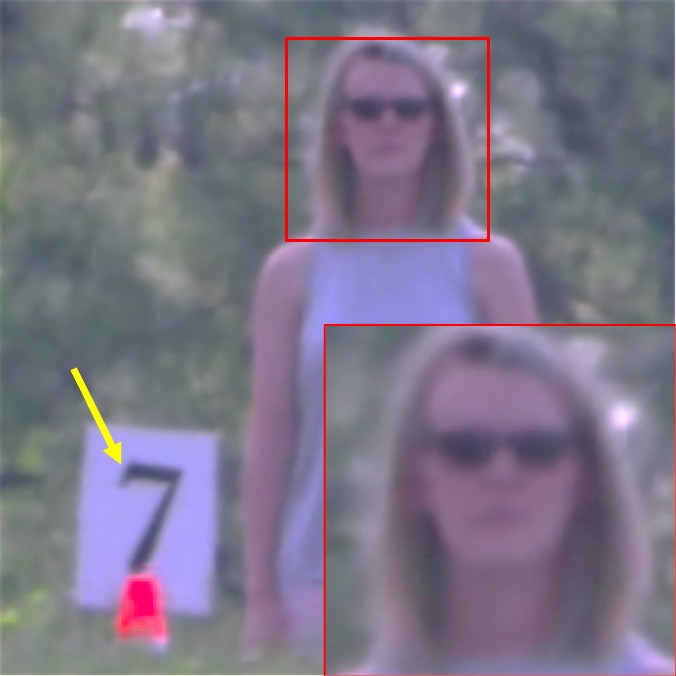}}
    \hfill
  \subfloat[ATNet \cite{Nair_2021_a}]{%
      \includegraphics[width=0.14\linewidth]{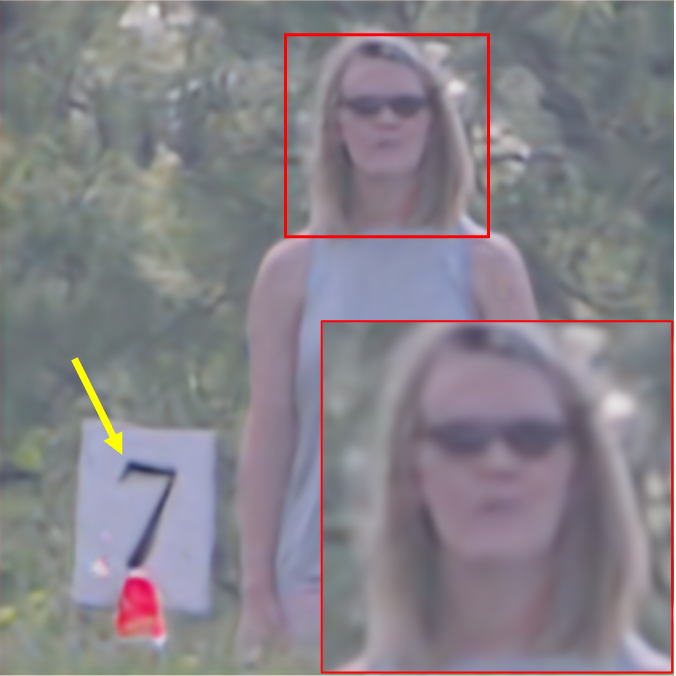}}
    \hfill
  \subfloat[AT-DDPM \cite{nair2023ddpm}]{%
      \includegraphics[width=0.14\linewidth]{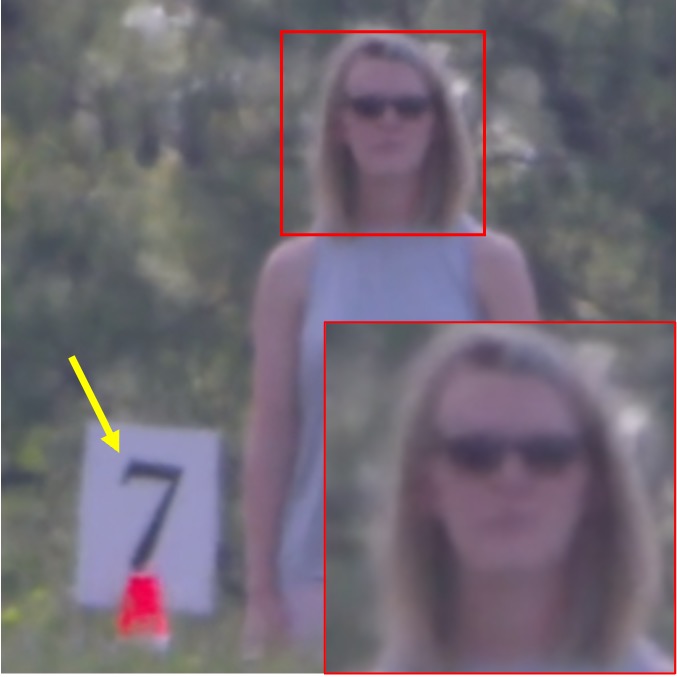}}
    \hfill
  \subfloat[PiRN-SR \cite{Jaiswal_2023_ICCV}]{%
      \includegraphics[width=0.14\linewidth]{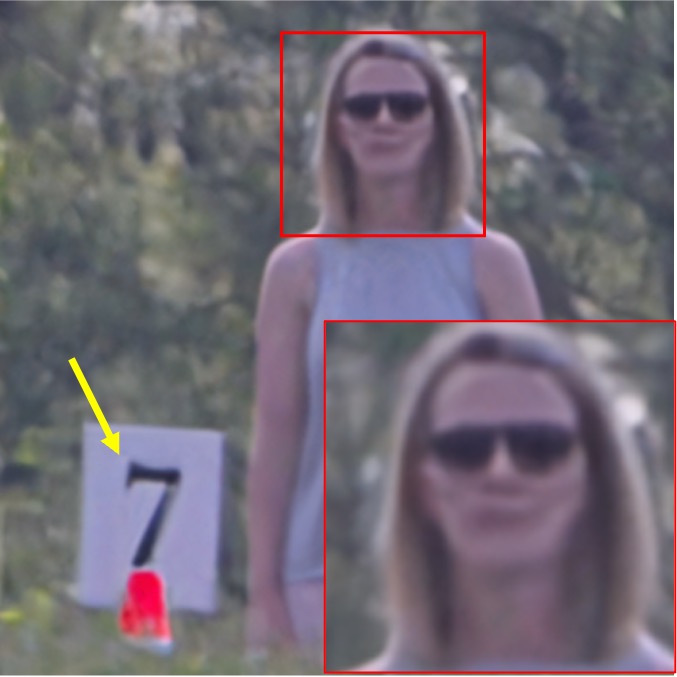}}
    \hfill

  \subfloat[\textbf{DATUM* [Ours]}]{%
      \includegraphics[width=0.14\linewidth]{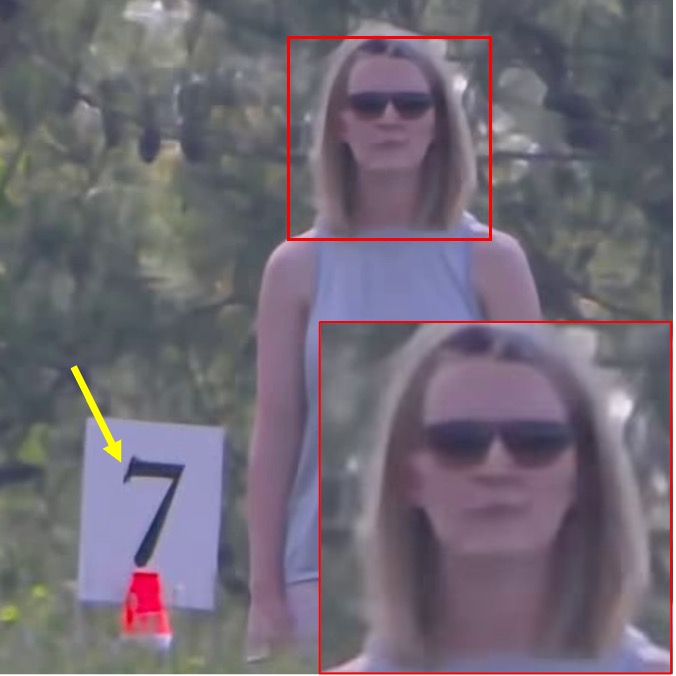}}
    \hfill
  \subfloat[TSRWGAN* \cite{Jin_2021_a}]{%
    \includegraphics[width=0.14\linewidth]{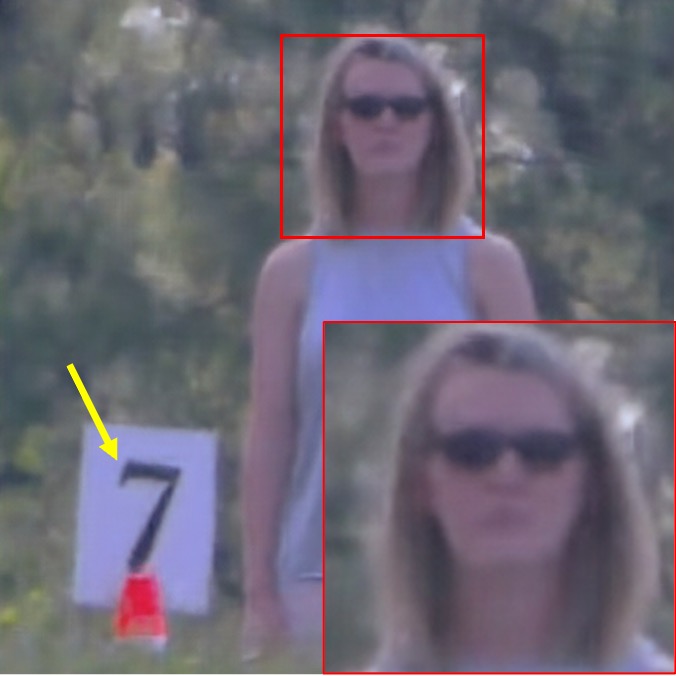}}
   \hfill
  \subfloat[TMT* \cite{Zhang_2022_a}]{%
        \includegraphics[width=0.14\linewidth]{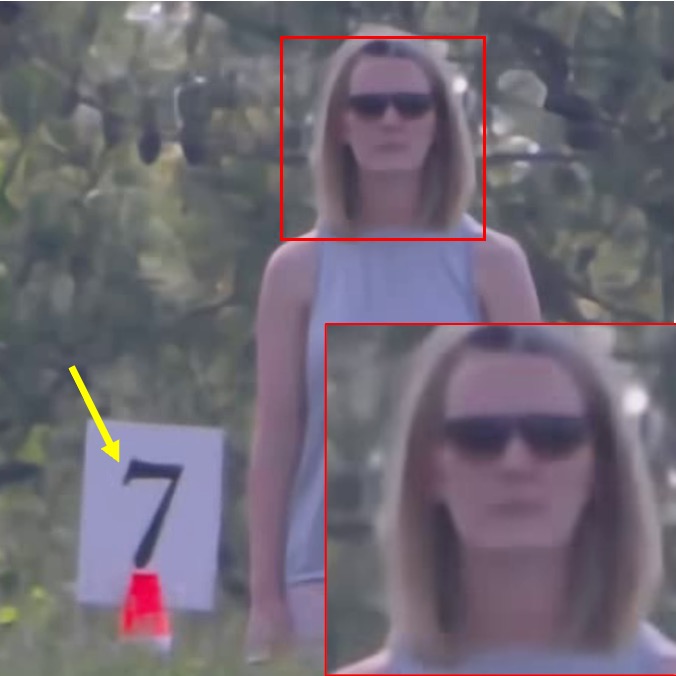}}
    \hfill
  \subfloat[RNN-MBP* \cite{Mao_2022_a}]{%
        \includegraphics[width=0.14\linewidth]{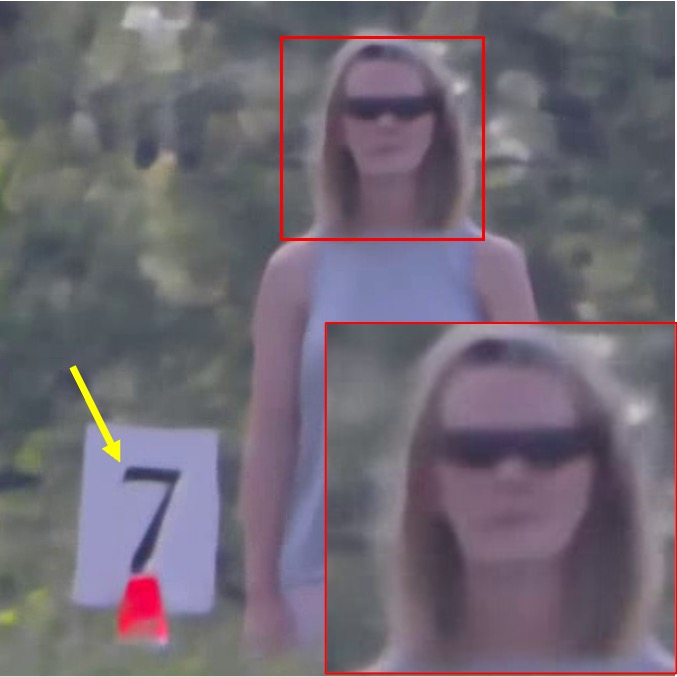}}
    \hfill
  \subfloat[RVRT* \cite{liang2022rvrt}]{%
      \includegraphics[width=0.14\linewidth]{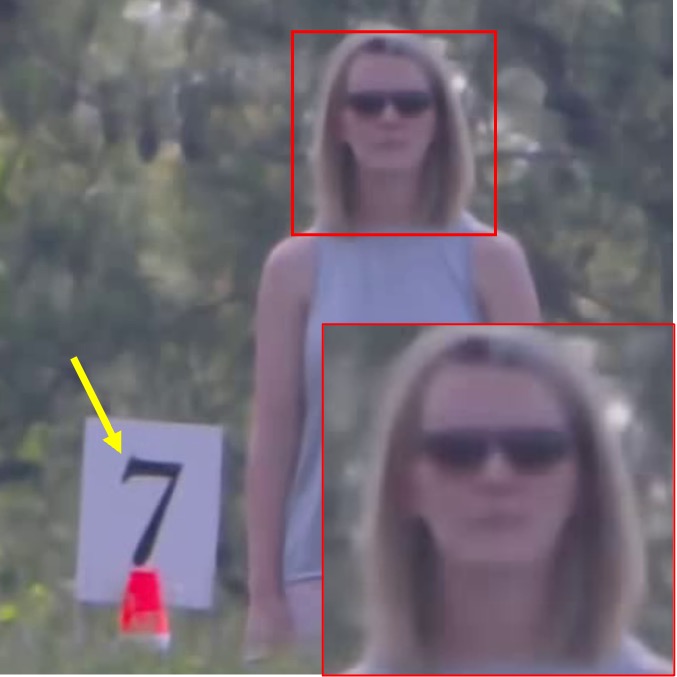}}
    \hfill
  \subfloat[VRT* \cite{Jin_2021_a}]{%
      \includegraphics[width=0.14\linewidth]{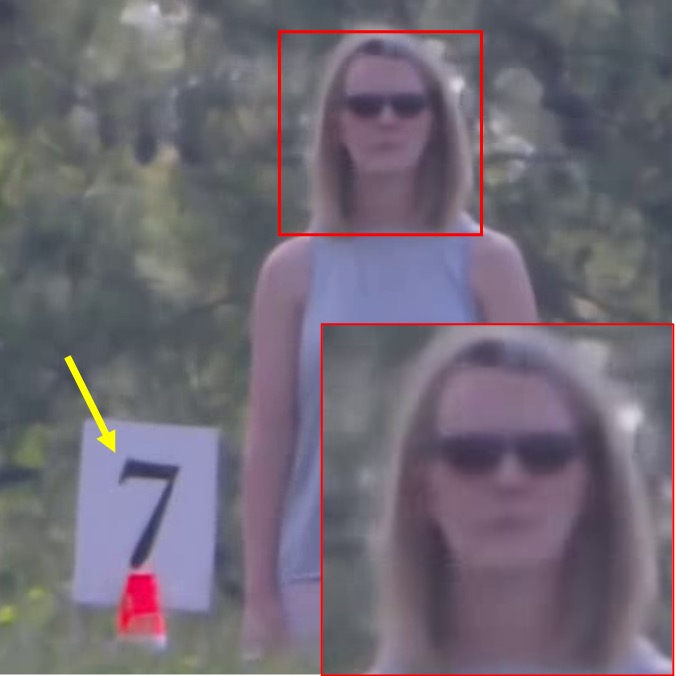}}
    \hfill
  \subfloat[ESTRNN* \cite{zhong2022real}]{%
      \includegraphics[width=0.14\linewidth]{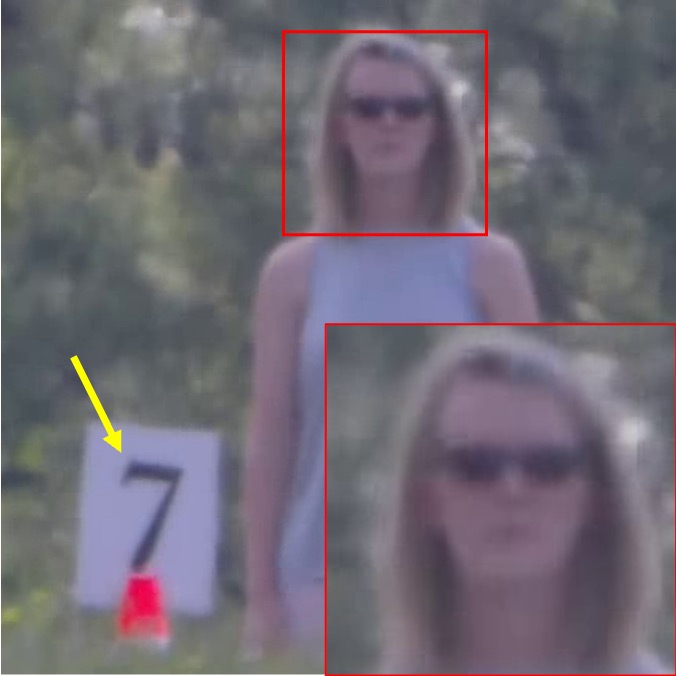}}
    \hfill
  \caption{Qualitative comparison on a dynamic scene sample from the BRIAR dataset \cite{cornett2023expanding}. Figures on the top row are the original restoration results of corresponding TM methods. Figures on the bottom row are models (marked by *) trained on \emph{ATSyn-dynamic} dataset.}
  \label{fig:briar} 
\end{figure*}

Given the impracticality of directly obtaining ground truth images for real-world turbulence scenarios, quantitative performance evaluation typically involves applying restored images to downstream tasks, as noted in \cite{Mao_2022_a, Jaiswal_2023_ICCV, nair2023ddpm}. Adopting this approach, we evaluated various restoration methods using the turbulence text dataset. The results are presented in Fig. \ref{fig:quant_text}, revealing two key insights: 1) our proposed ATSyn-static dataset enhances the generalization capabilities of other TM methods. 2) on both synthetic and real-world sequences, DATUM consistently outperforms other models trained on our dataset. To further validate the effectiveness of our modifications to the Zernike-based simulator, we extensively compared DATUM trained on our ATSyn-dynamic dataset and TMT's dataset \cite{Zhang_2022_a}. We first enhance the long-range subset in the BRIAR dataset \cite{cornett2023expanding} by those two versions, run the same pre-trained face recognition model \cite{Kim_2022_CVPR} on the enhanced images, and it yields the result provided in Table \ref{tab:face}. We can observe the ATSyn-dynamic dataset improved network performance on real-world videos compared to the \cite{Zhang_2022_a} dataset. These comparisons demonstrate our method facilitates better generalization of both scene types than other existing datasets.

We also provide a qualitative comparison in Fig. \ref{fig:text} and \ref{fig:briar} to demonstrate the advance of our network and dataset. By comparing the same networks trained by our data and their original checkpoints, our data enhances their generalization capability. On the other hand, by comparing all networks trained on our dataset, our model significantly outperforms other networks.
\section{Conclusion}

In this research, we introduced a novel approach leveraging deep learning to address the enduring challenge of atmospheric turbulence mitigation. Taking a translational perspective, our method integrated the strengths of traditional turbulence mitigation (TM) techniques into a neural network architecture. This fusion elevated our network to state-of-the-art performance while ensuring significantly enhanced efficiency and speed compared to prior TM models. Additionally, we developed a physics-based synthesis method that accurately models the degradation process. This led to the creation of an extensive synthetic dataset covering a diverse spectrum of turbulence effects. Utilizing this dataset, we facilitated a stronger generalization capability for data-driven models than other existing datasets.

\noindent \textbf{Acknowledgments and Disclosure of Funding}
The research is based upon work supported in part by the Intelligence Advanced Research Projects Activity (IARPA) under Contract No. 2022‐21102100004, and in part by the National Science Foundation under the grants CCSS-2030570 and IIS-2133032. The views and conclusions contained herein are those of the authors and should not be interpreted as necessarily representing the official policies, either expressed or implied, of IARPA, or the U.S. Government. The U.S. Government is authorized to reproduce and distribute reprints for governmental purposes notwithstanding any copyright annotation therein. 



\begin{appendices}

\vspace{2em}

\renewcommand{\appendixname}{Appendix}

\centerline{\Large\bfseries \appendixname}
\vspace{1em}

\section{Additional Experiments}
\label{sec:add_exp}


\subsection{Visualization of flow refinement in DAAB}
The Deformable Attention Alignment Block (DAAB) is designed to align features from a current time frame, denoted as time $t$, with reference features from a preceding frame, time $t-1$, during forward temporal propagation. This approach differs fundamentally from traditional optical flow methods, which align two degraded frames between times $t$ and $t-1$ by $O^f_{t\rightarrow t-1}$. DAAB instead aligns the feature map of the current frame $t$ with a potentially tilt-corrected reference feature from the previous frame $t-1$. The effectiveness of DAAB has been substantiated in previous ablation studies.

\begin{figure*}[t]
    \centering
    \footnotesize
    \begin{tabular}{ccc}
        \subfloat[input frame $t$]{%
        \includegraphics[width=0.31\linewidth]{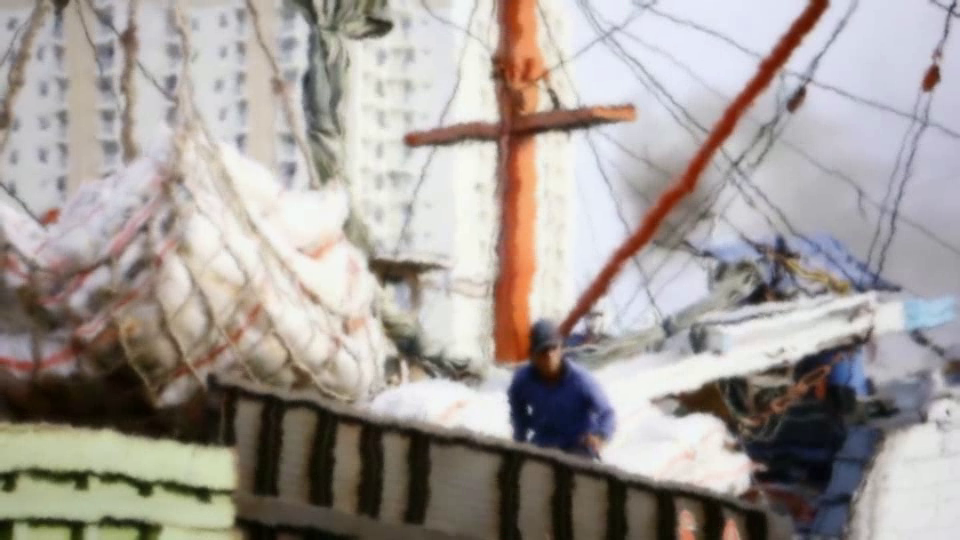}} \hfill 
        \subfloat[input frame $t-1$]{%
        \includegraphics[width=0.31\linewidth]{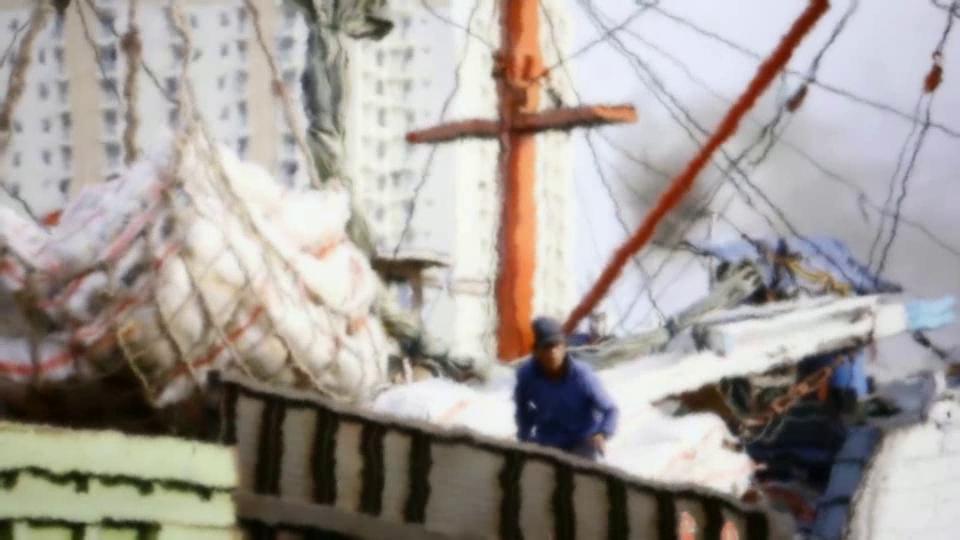}} \hfill
        \subfloat[restored frame $t$ by DATUM]{%
        \includegraphics[width=0.31\linewidth]{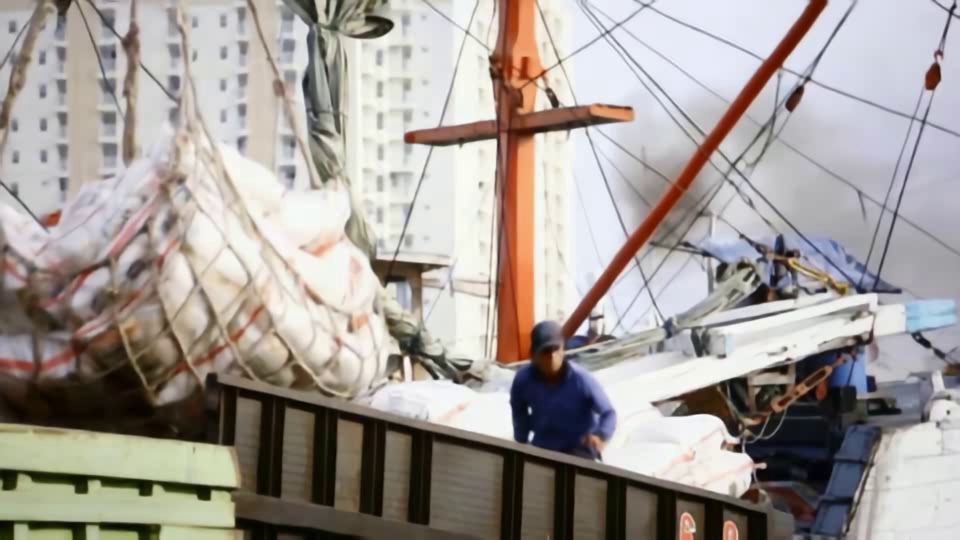}}\\    
        \subfloat[optical flow from $t$ to $t-1$ $O^f_{t\rightarrow t-1}$]{%
        \includegraphics[width=0.31\linewidth]{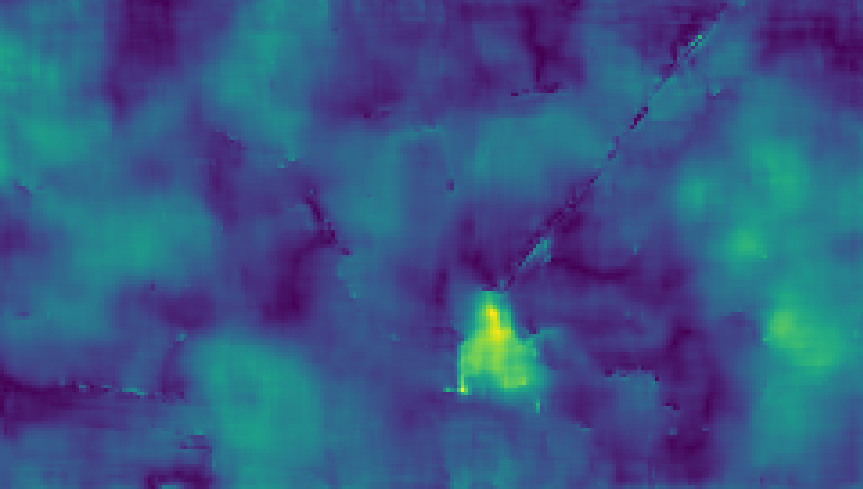}} \hfill
        \subfloat[refined feature to reference flow $O^{f\rightarrow r}_{t}$]{%
        \includegraphics[width=0.31\linewidth]{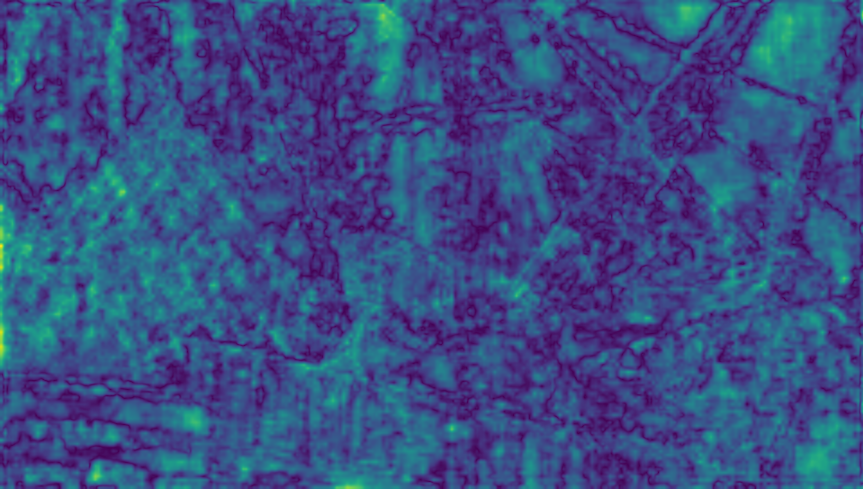} } \hfill
        \subfloat[estimated inverse tilt field $\hat{\mathcal{T}}^{-1}_{t}$]{%
        \includegraphics[width=0.31\linewidth]{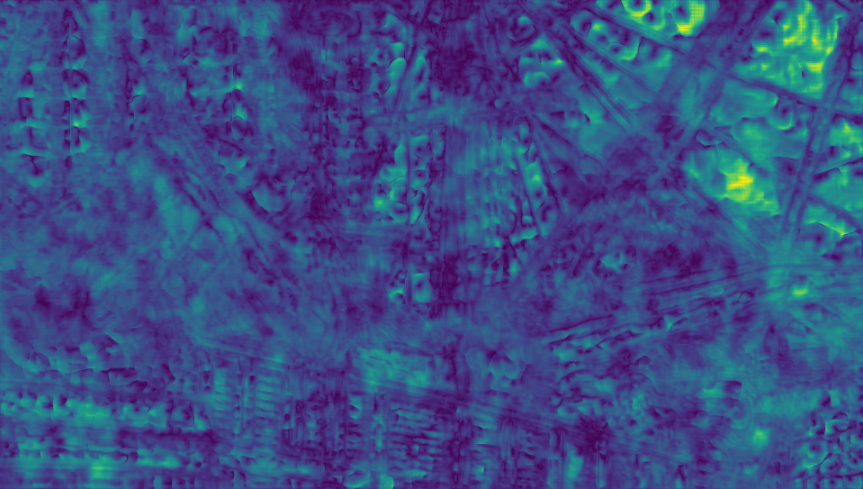} }
    \end{tabular}
    \caption{Visualization of the flow refinement for feature-reference registration in DAAB. (d), (e) and (f) show the magnitude of the associated deformation field. We ignore the directional information because it is relatively random. Note both (d) and (e) are measured in $1/4$ resolution, while (f) is in full resolution, which aims to register shallow features extracted from (a) to those from (c).}
    \label{fig:DAAB_flow}
\end{figure*}

To further illustrate its efficacy, we provide an additional visualization in Fig. \ref{fig:DAAB_flow}, leading to several critical observations:

\begin{enumerate}
\item The original flow estimation $O^f_{t\rightarrow t-1}$ captures mild motion, such as that of a person, but introduces noise due to random pixel displacements in static image regions.
\item The refined flow that registers $f_{t}$ to $r_{t-1}$ is more dependent on the structural information and less sensitive to the mild motion.
\item The magnitude of the refined flow under DAAB exhibits a pattern indicative of tilt rectification.
\item Additional visualization of the estimated reverse tilt field $\hat{\mathcal{T}}^{-1}_{t}$, which adjusts frame $t$ to a tilt-free state, demonstrates that $O^{f\rightarrow r}_{t}$ aligns more closely with $\hat{\mathcal{T}}^{-1}_{t}$. This alignment is in line with the intended design of DAAB for effective feature-reference registration.
\end{enumerate}

\subsection{More qualitative comparisons on real-world image sequences}

\begin{figure}[t]
    \centering
    \footnotesize
    \begin{tabular}{cccccc}
        \subfloat[Text image input]{%
        \includegraphics[width=0.31\linewidth]{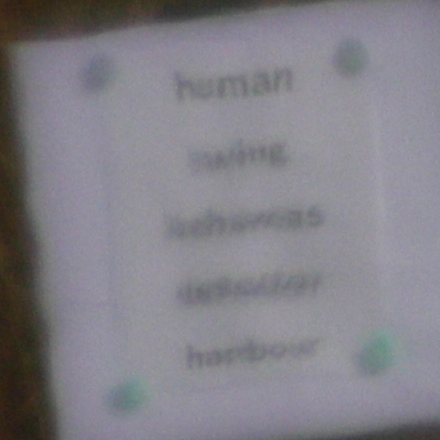}} \hfill
        \subfloat[Restored by ATNet]{%
        \includegraphics[width=0.31\linewidth]{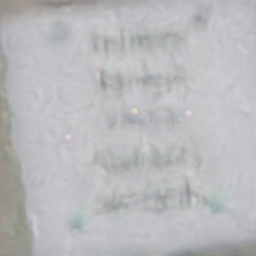}} \hfill
        \subfloat[Restored by DATUM]{%
        \includegraphics[width=0.31\linewidth]{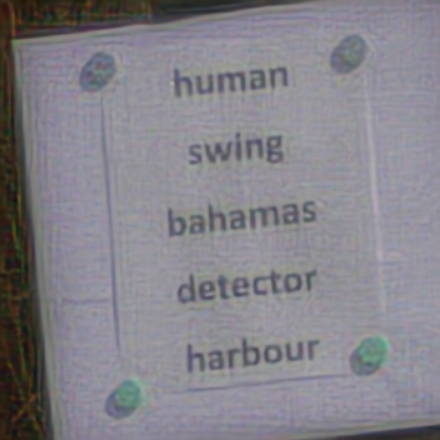}} \\ 
        \subfloat[OTIS input]{%
        \includegraphics[width=0.31\linewidth]{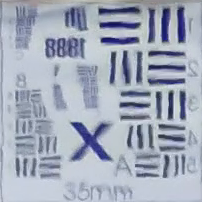}} \hfill
        \subfloat[Restored by ATNet]{%
        \includegraphics[width=0.31\linewidth]{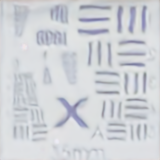} } \hfill
        \subfloat[Restored by DATUM]{%
        \includegraphics[width=0.31\linewidth]{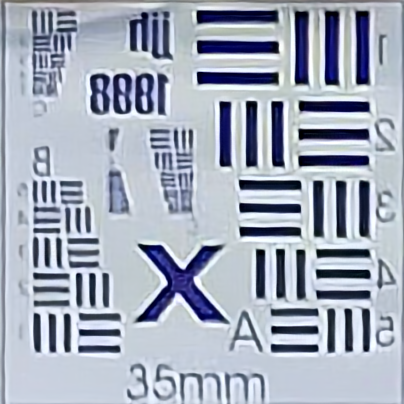}}
    \end{tabular}
    \caption{Cases of ATNet \cite{Nair_2021_a} restoration on real-world static scene images. The text image is the 49th frame of the 94th sequence in \cite{UG2}, and the OTIS image is the 24th frame of the pattern 13 from the \cite{Gilles_2017_a} dataset.}
    \label{fig:atnet}
\end{figure}

\noindent \textbf{ATNet \cite{Nair_2021_a} on the static scene data}. In Fig. 6, we show the restoration results of NDIR \cite{Li_2021_a} rather than the ATNet \cite{Nair_2021_a}. NDIR is an unsupervised multi-frame pixel alignment network without a deblurring function, while ATNet is a single-frame-based general TM network. However, ATNet's inference is not successful. The results on some static scene data are shown in Fig. \ref{fig:atnet}, which suggests it is challenging for this single-frame-based model to deal with medium to strong turbulence, while our methods can handle much wider turbulence conditions. 

\begin{figure*}[t]
    \centering
    \footnotesize
    \begin{tabular}{ccc}
        \includegraphics[width=0.31\linewidth]{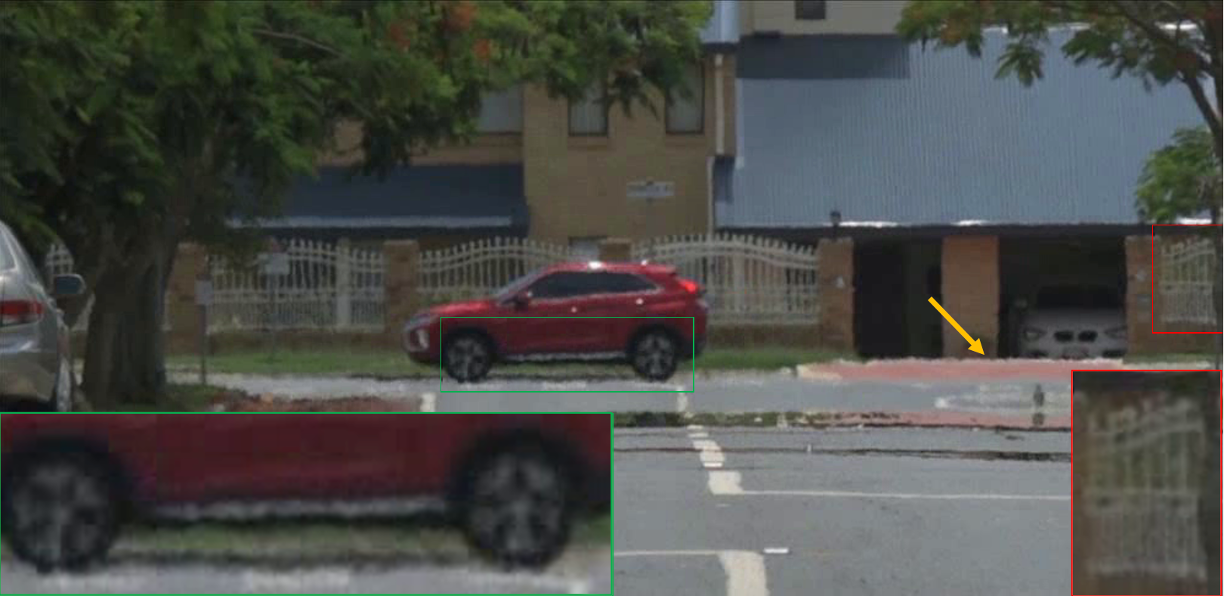} \hfill 
        \includegraphics[width=0.31\linewidth]{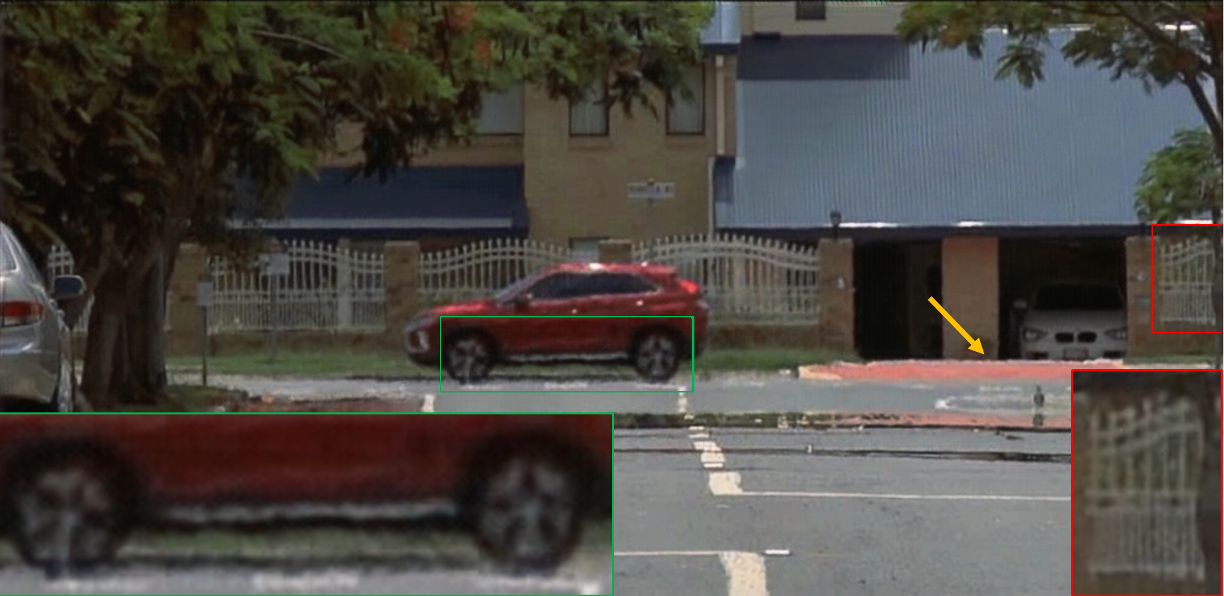} \hfill
        \includegraphics[width=0.31\linewidth]{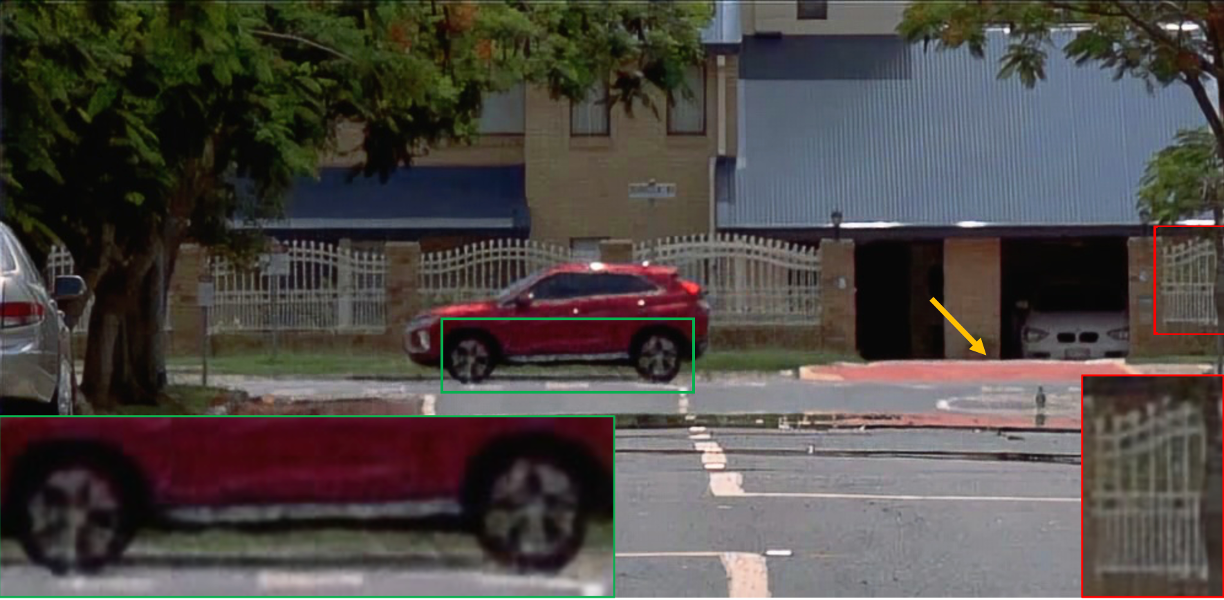}\\
        \includegraphics[width=0.31\linewidth]{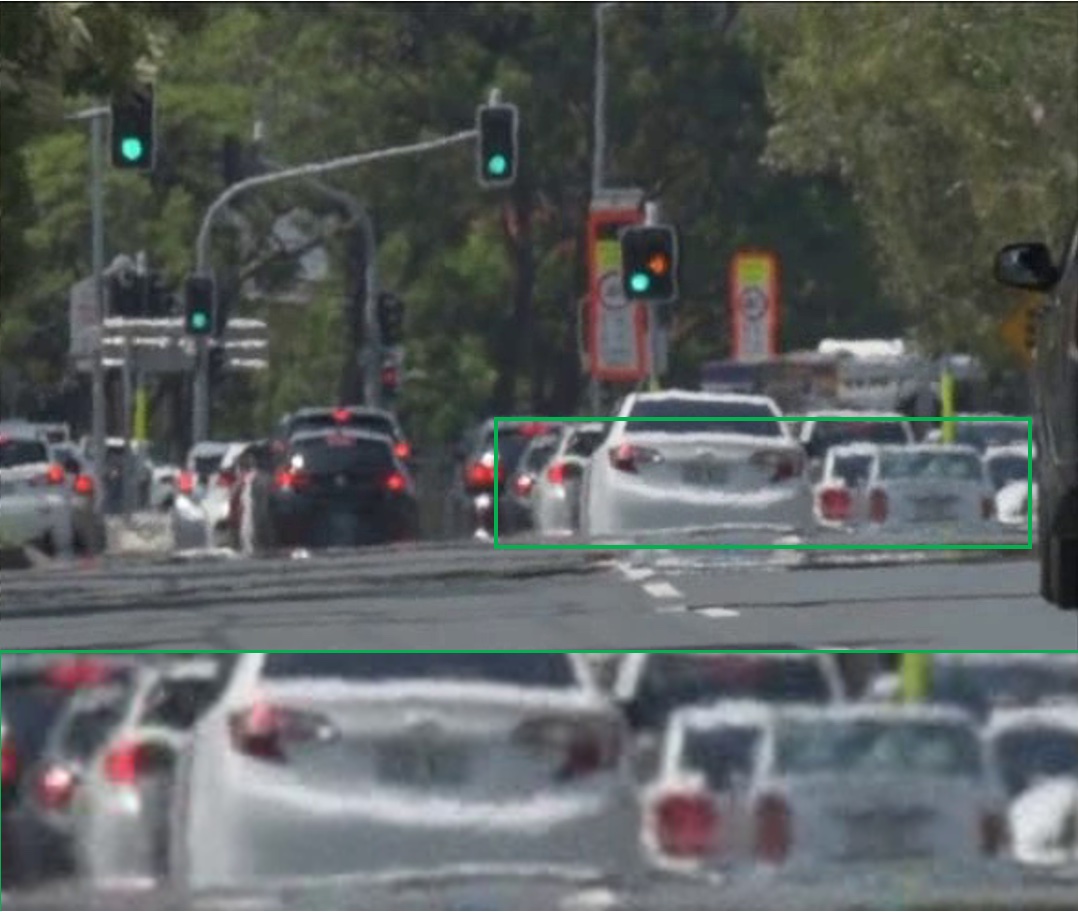} \hfill 
        \includegraphics[width=0.31\linewidth]{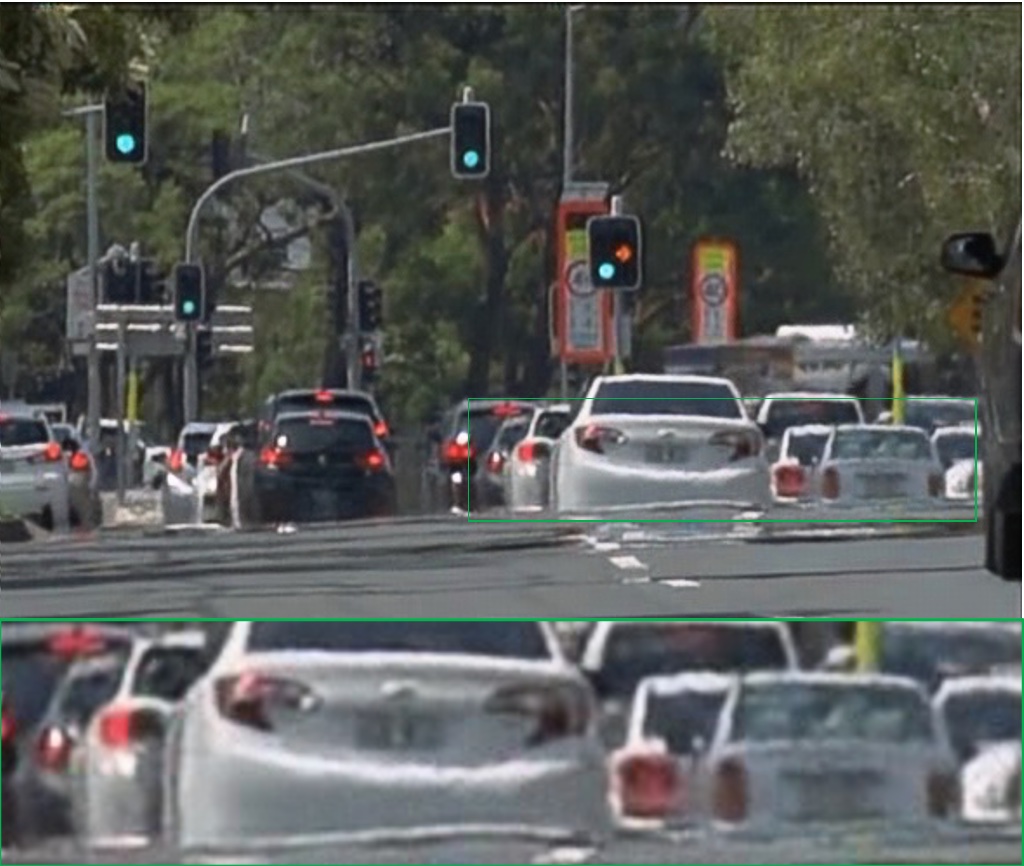} \hfill
        \includegraphics[width=0.31\linewidth]{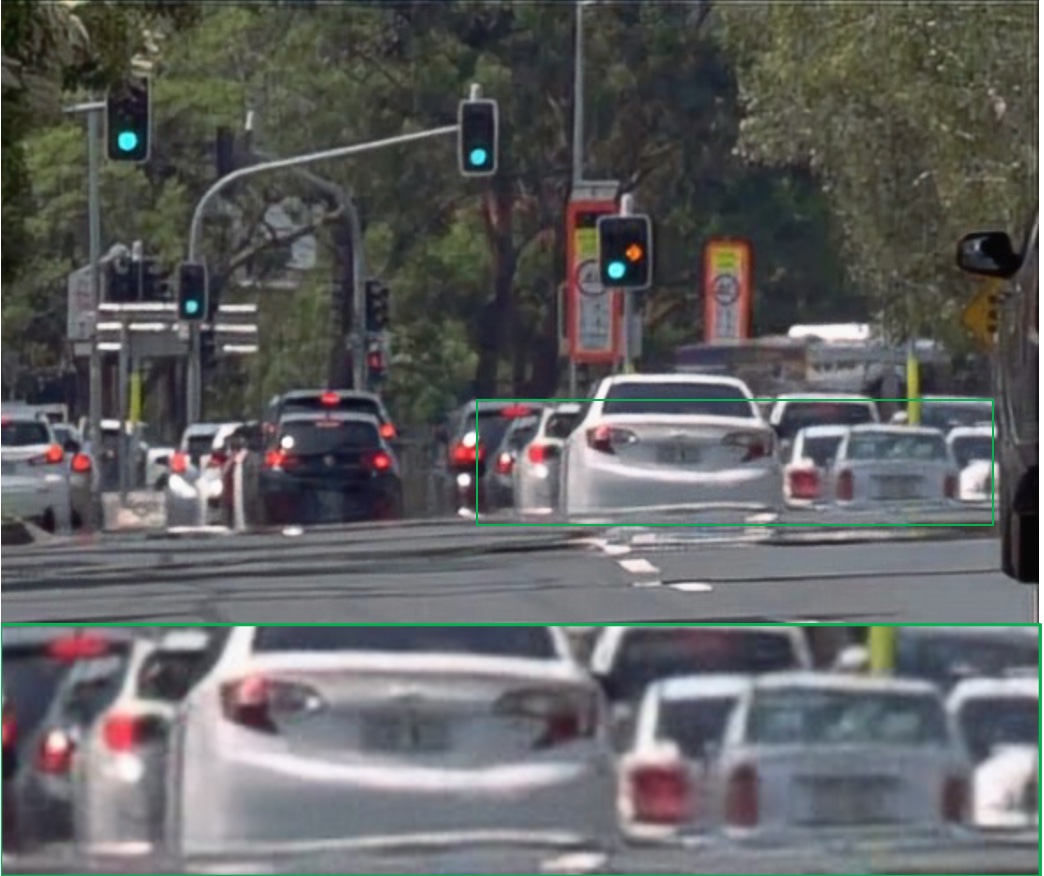}\\    
        \subfloat[input]{%
        \includegraphics[width=0.31\linewidth]{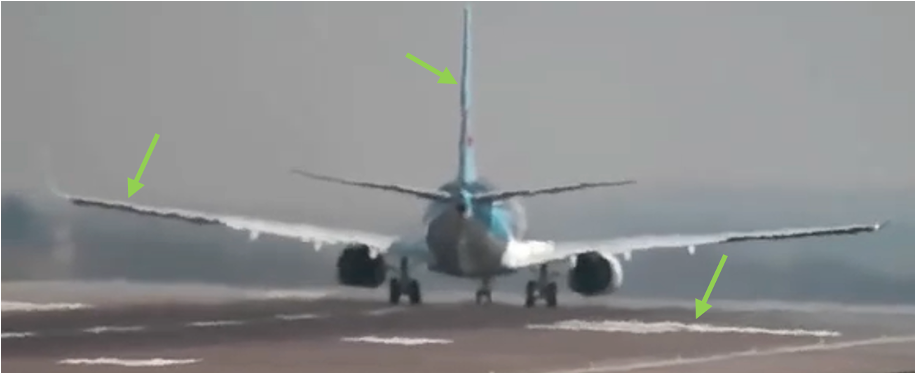}} \hfill
        \subfloat[Original TSRWGAN \cite{Jin_2021_a}]{%
        \includegraphics[width=0.31\linewidth]{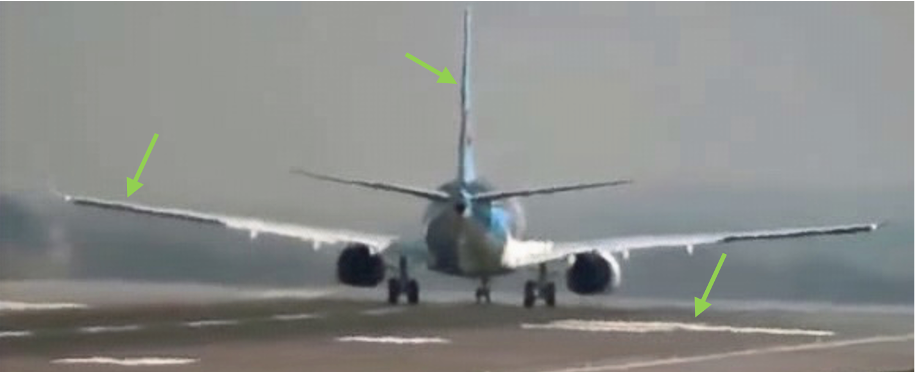} } \hfill
        \subfloat[Fine-tuned TSRWGAN]{%
        \includegraphics[width=0.31\linewidth]{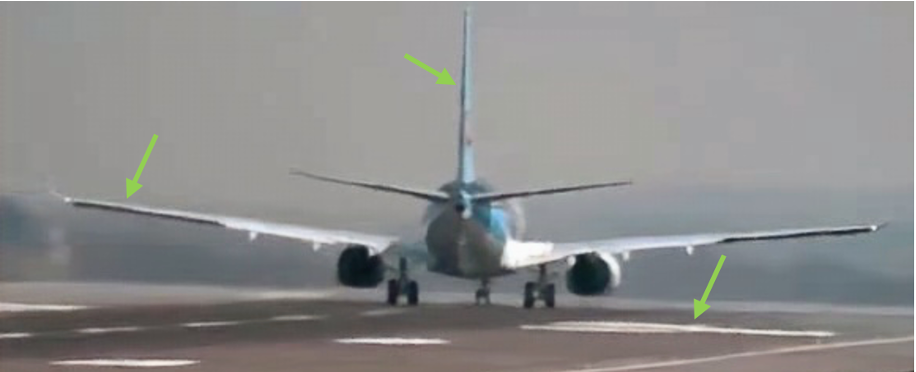}}
    \end{tabular}
    \vspace{-1em}
    \caption{Compare the TSRWGAN \cite{Jin_2021_a} trained on the original dataset and our dataset, the first two rows are real-world samples from \cite{Jin_2021_a}'s dataset, and the bottom row is from \cite{Anantrasirichai_2022_a}'s real-world videos. In column (c), we present the fine-tuned TSRWGAN on our ATSyn-dynamic dataset. From the comparison, it's easy to conclude that our ATSyn dataset helps the previous turbulence mitigation network generalize better on their own testing videos and other samples.}
    \label{fig:tsrwgan}
\end{figure*}

\textbf{Compare with TSRWGAN \cite{Jin_2021_a}} We address the generalization facilitated by our data synthesis method. A qualitative comparison was made between the original TSRWGAN \cite{Jin_2021_a} and our fine-tuned version on \cite{Jin_2021_a}'s real-world dynamic scenes along with a cross-dataset evaluation between these two versions on \cite{Anantrasirichai_2022_a}'s real-world dynamic scenes. The result is shown in Figure \ref{fig:tsrwgan}. The original model shows a limit in generalization when adapting to a different dataset, but our fine-tuned version is more generalizable due to ATsyn's wide range of turbulent conditions. The original TSRWGAN model is trained from the simulator from \cite{repasi2011computer} and physical simulation by heating the air along a relatively short path. Their numerical simulator can generate physics-based tilt and spatially varying blur, but higher-order aberrations are not modeled. Their physical simulator tends to generate spatially highly correlated distortion but a weak blurry effect. Because of these limitations in their generation, their generalization to other datasets suffers as a result.

\begin{figure*}[t]
    \centering
    \footnotesize
    \begin{tabular}{ccc}
        \includegraphics[width=0.31\linewidth]{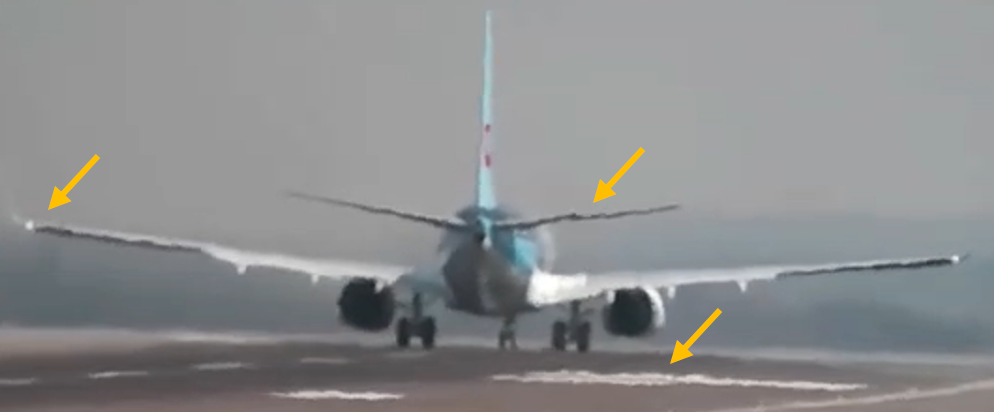} \hfill 
        \includegraphics[width=0.31\linewidth]{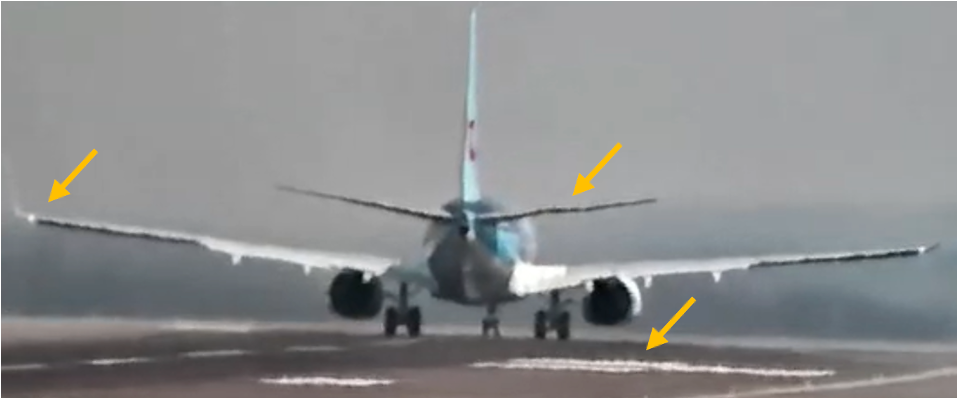} \hfill
        \includegraphics[width=0.31\linewidth]{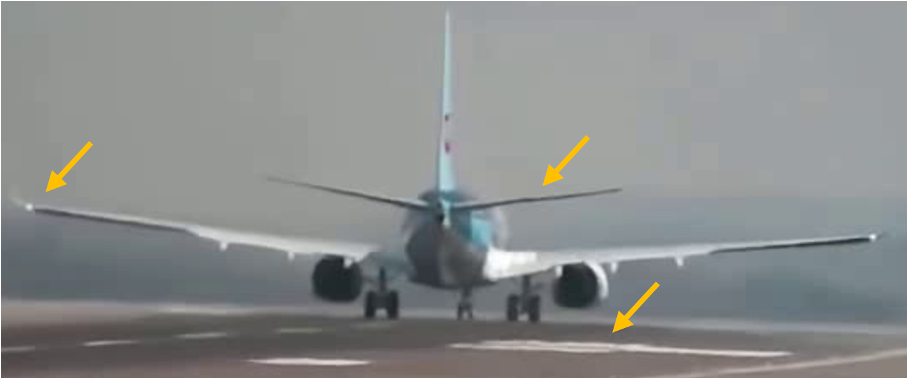}\\    
        \subfloat[input]{%
        \includegraphics[width=0.31\linewidth]{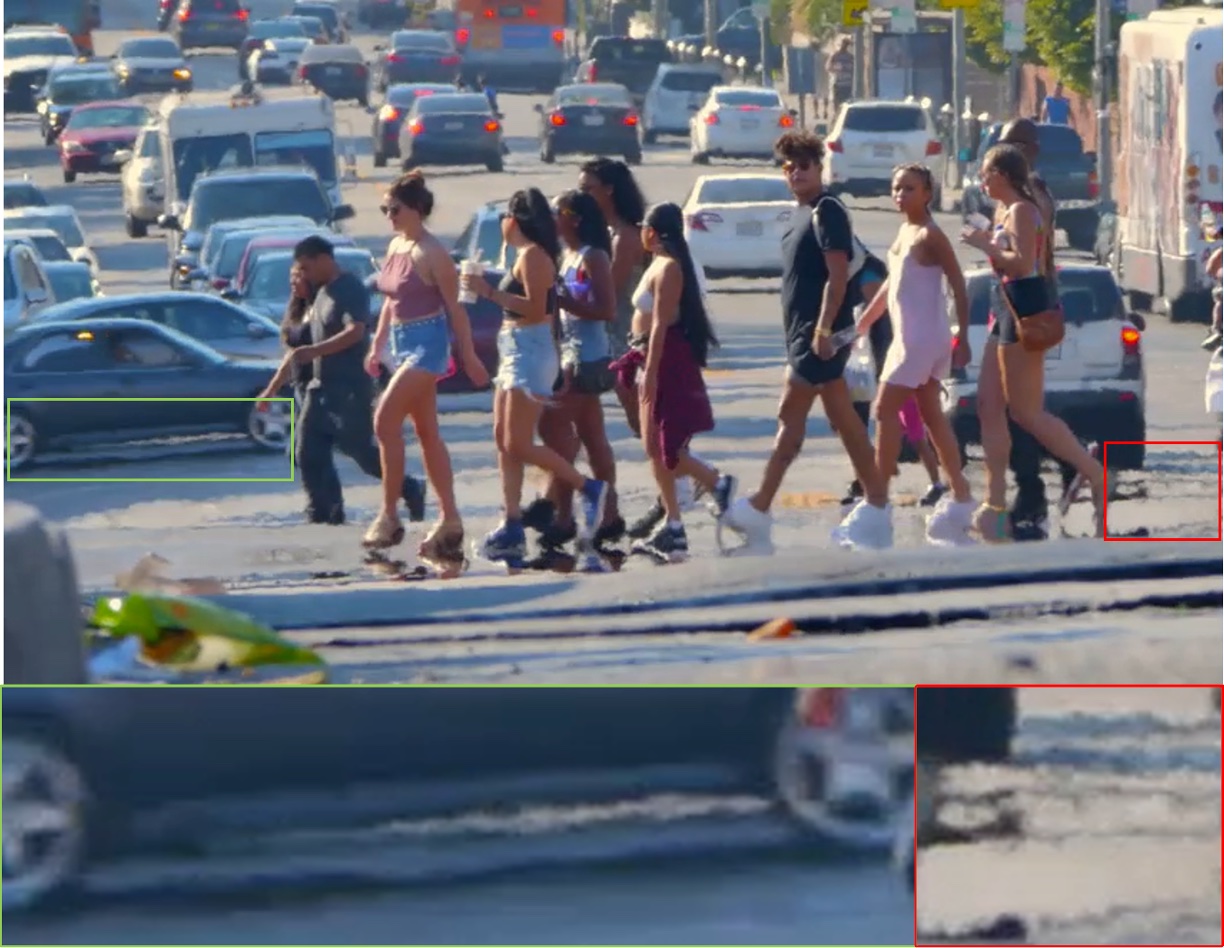}} \hfill
        \subfloat[Restored by Complex-CNN \cite{Anantrasirichai_2022_a}]{%
        \includegraphics[width=0.31\linewidth]{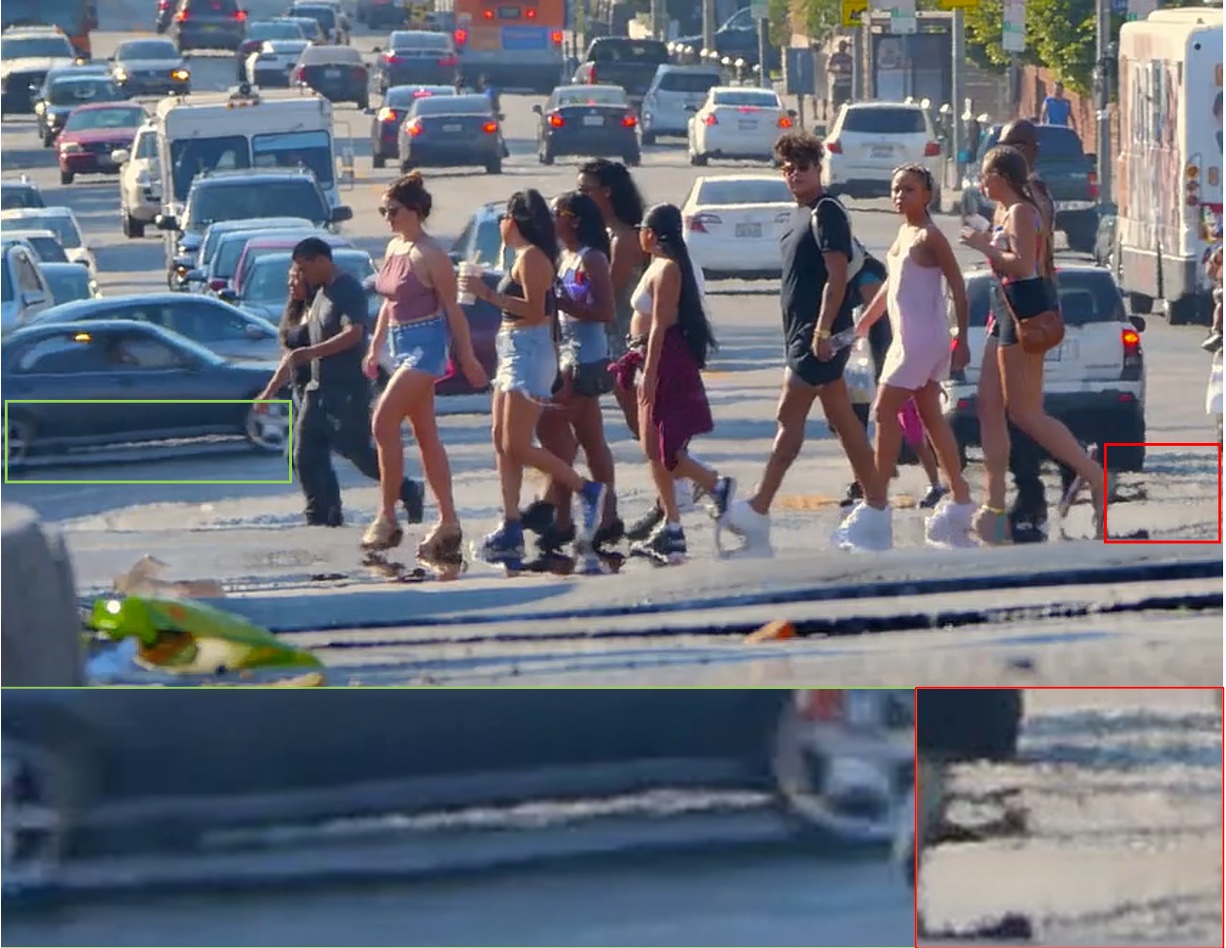} } \hfill
        \subfloat[Restored by DATUM]{%
        \includegraphics[width=0.31\linewidth]{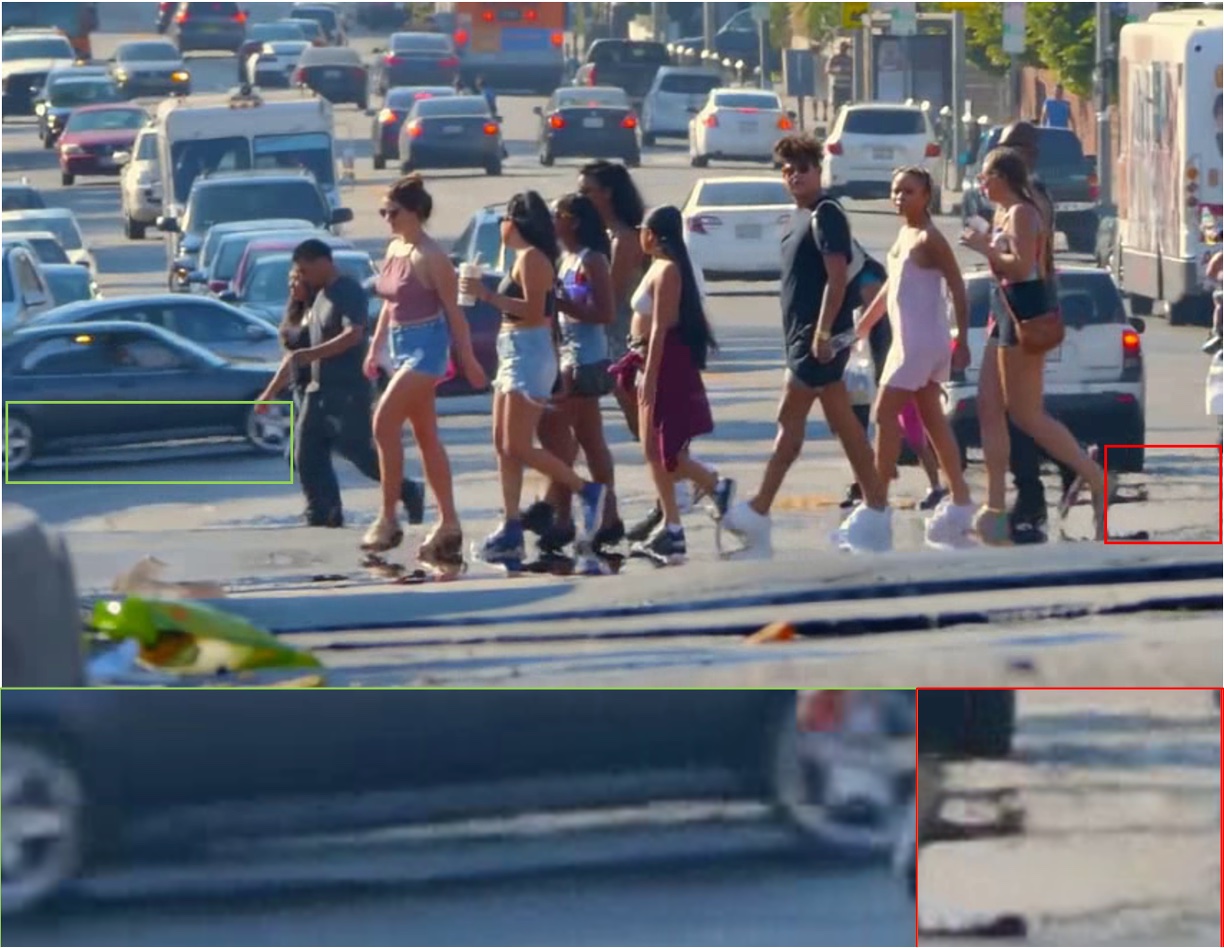}}
    \end{tabular}
    \vspace{-1em}
    \caption{Comparison with \cite{Anantrasirichai_2022_a} on their real-world dataset, zoom in for a better view.}
    \label{fig:complex_cnn}
\end{figure*}

\noindent \textbf{Compare with Complex-CNN \cite{Anantrasirichai_2022_a}}
A complex-valued convolutional neural network (CNN) \cite{Anantrasirichai_2022_a} was proposed to remove turbulence-related degradation from videos. Their synthetic training data comes from a simulator that models the tilt and blur via a low-order approximation, with the blur kernel being sampled from 9 given point spread functions. Without access to their trained model, we cannot fine-tune. However, with some results available, we may compare the performance of our restored videos with theirs on their dataset. We provide this comparison in Figure \ref{fig:complex_cnn}.

\begin{figure*}[h]
    \centering
    \includegraphics[width = 0.99\linewidth]{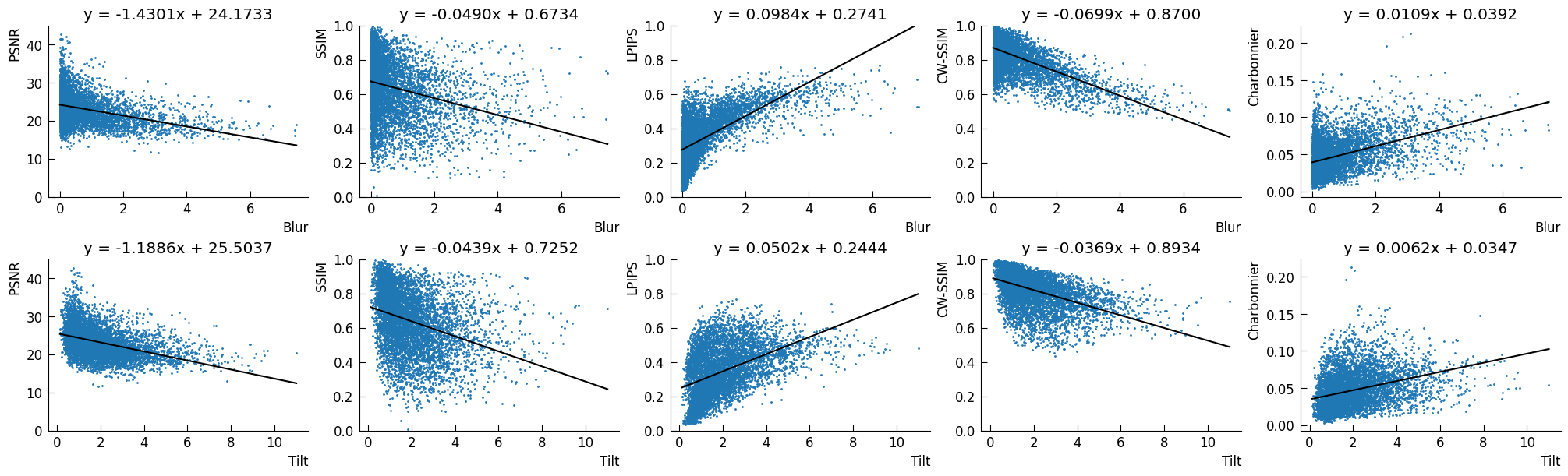}
    \caption{Image quality metrics. The x-axis is the score of blur or tilt; y-axis is the image quality score measuring the degradation with respect to the clean image. We measured PSNR, SSIM, LPIPS, CW-SSIM, and the Charbonnier score, which serves as the loss of our optimization for turbulence mitigation.}
    \label{fig:metrics}
\end{figure*}

\subsection{Image quality metrics for turbulence mitigation}
\label{metrics}
 In our empirical study, we observed a high correlation between two commonly used metrics: Peak Signal-to-Noise Ratio (PSNR) and Structural Similarity Index Measure (SSIM). Atmospheric turbulence typically induces blur and pixel displacement in images. While the blurring effect is readily noticeable in both human and computer vision applications, minor pixel displacements often remain less perceptible. However, PSNR and SSIM are particularly sensitive to pixel displacements. This sensitivity raises the need for additional metrics to enable a more comprehensive performance evaluation. We investigate the Complex-wavelet SSIM (CW-SSIM), a variant of SSIM that is less sensitive to mild pixel displacement, and the Learned Perceptual Image Patch Similarity (LPIPS) for this purpose.

With the turbulence simulator detailed in the section \ref{sec:zernike_simulator}, we can synthesize different levels of atmospheric turbulence. For the Zernike-based simulator, the turbulence effect can be quantified by the magnitude of Zernike coefficients, which indicate the properties of phase distortion caused by anisoplanatic turbulence. We compute different image quality scores for each pair of degraded and clean images. To assure the robustness of our analysis, we randomly chose 1000 images from the Places dataset \cite{zhou2017places} as clean images and simulated nine degraded samples for each, so we draw 9000 samples in total and show the relationship between the strength of turbulence degradation and image quality metrics in Fig. \ref{fig:metrics}. Note we separate the tilt and blur effects, although they are highly correlated. The score of tilt is the average magnitude of pixel displacement on an image, and the score of blur is calculated by 
\begin{equation*}
    \text{blur} = k_{b} \frac{\sum_{\vx}(\sqrt{\sum_{i=3:36}a^2_{\vx, i}})}{HW},
\end{equation*}
where $\vx = (x,y)$ is the pixel coordinate on each image, $H$, $W$ are the height and width of the image, and $k_{b}$ is the scaling factor determined by the relative size of blur kernels. 

From Fig. \ref{fig:metrics}, we can find the SSIM is less sensitive to turbulence degradation than the others, and CW-SSIM is more sensitive than LPIPS. Thus, we selected PSNR and CW-SSIM as our restoration quality estimators.

\section{Zernike-based Turbulence Simulator}
\label{sec:zernike_simulator}
\subsection{General Theory}

We adopt the model of the atmospheric degradations to be exclusively phase distortions, which can be represented via the Zernike polynomials $\{\mZ_i\}$ as a basis, with coefficients $\va_{\vx, i}$ \cite{Noll_1976_a, Chimitt_2020_a}. We set $i\! \in \! \{1,2,3,\cdots,36\}$ with $\mZ_{\{2,3\}}$ influencing the pixel displacement $\boldsymbol{\mathcal{T}}$ and higher order coefficients $\mZ_{\{i \geq 4\}}$ forming the blurry effect $\boldsymbol{\mathcal{B}}$ in the image plane. With this, the kernel of $\boldsymbol{\mathcal{B}}_{\vx}$ can be written as:
\begin{equation}
    \boldsymbol{\mathcal{B}}_{\vx} \approx \left| \mathcal{F} \left\{ \text{exp}\left(-j\sum_{i=4}^{36} a_{\vx,i} \mZ_i \right) \right\} \right|^2,
    \label{eq: Fourier_incoherent_PSF}
\end{equation}
where $\mathcal{F}$ denotes the Fourier transform. Adopting the wide sense stationary model for the Zernike coefficients \cite{Chimitt_2020_a, chimitt2022real}, one can generate $\va_{\vx, i}$ in parallel by Fourier Transform. It is worth noting $\va_{\vx, \{2,3\}}$ can be excluded here as they contribute the pixel-shifting $\boldsymbol{\mathcal{T}}$, and thus may be separated according to \cite{Chan_2022_a}.

Hence, the phase distortions caused by atmospheric turbulence can be further described by a random vector $\va_\vx = [a_{\vx,1}, a_{\vx,2}, a_{\vx,3}, \ldots]^T$ at each pixel $\vx$ in an image, which forms a set of random fields \cite{chimitt2022real}. As stated by Noll \cite{Noll_1976_a}, each vector is a 0-mean Gaussian vector with a specified covariance matrix, 
\begin{equation}
    \E[\va_\vx \va_\vx^T] = \mR.
\end{equation}
Noll used the Zernike polynomials to describe the phase distortions resulting from a point source, resulting in the basis representation:
\begin{equation}
    \phi_\vx(R\vrho) = \sum_i a_{\vx,i} \mZ_i(\vrho),
\end{equation}
where $\vrho$ is a vector defined over the unit circle, and $R$ is the radius of the imaging system's aperture. 

This concept has been generalized to include \emph{separate} positions $\vx$ and $\vx'$, which form a covariance tensor $\E[a_{\vx,i} a_{\vx',j}]$. \cite{chimitt2022real} states that one may quickly generate the turbulent distortions for an image of size $H \times W$, within suitable approximation, from these components in the following way:

\begin{enumerate}
    \item For $i \in \{1, 2, \ldots, 36\}$, compute the power spectral density (PSD) $\mS_i$ for each covariance function through the use of the Wiener–Khinchin theorem, $\boldsymbol{S}_i=\mathcal{F}\{ \E[a_{\vx,i} a_{\vx',j}] \}$, where $\mathcal{F}$ denotes the Fourier transform.
    
    \item Generate 36 zero-mean unit variance random fields according to the covariance function $\E[a_{\vx,i} a_{\vx',j}]$. This is done according to FFT-based methods, which use a complex white noise seed $\vn$ to form a field $\vv_i$ in the following way: $\vv_i = \text{real}(\mathcal{F}^{-1} \{ \sqrt{\mS_i} \vn \} )$.
    
    
    \item Perform a Cholesky decomposition of the matrix derived by Noll $\mR = \mL \mL^t$, which in our case is of size $36 \! \times \! 36$. Denoting the concatenated fields as $\vv = [\vv_{1}, \vv_{2}, \ldots, \vv_{36}]^T$ with dimensions $ 36\! \times\! H\! \times \! W$, the final output random fields may be generated as $\va' = \mL \vv$.
    
    \item Provide the Zernike coefficient fields $\va'$ to the Phase-to-Space transformation (P2S) to compute the PSF-basis coefficients $\vbeta_{\vx} = \mathcal{P}(\va'_{\vx, \{i \geq 4\}})$.
    
    \item Apply the image warping followed by the spatially varying blur by the P2S coefficients as described in the main body of the paper.
\end{enumerate}
For color images, the same process is carried out, with the spatially varying convolution occurring in the same way for all color channels in accordance with \cite{Mao_2021_a}. 

Although from a high level, the simulation process in this work is identical to that of \cite{chimitt2022real}, there are some critical differences:
\begin{enumerate}
    \item The spatially varying convolution is modified to match the image formation process more accurately. Though this is detailed in the paper, we provide additional evidence of the importance of this modification in a section \ref{reformulation}. This affects step (5) of the simulation.

    \item We use a reformulated expression $\E[a_{\vx,i} a_{\vx',j}]$ according to \cite{Chimitt_2023_a}, which we detail in the next two subsections. This reformulation leads to an exact solution rather than the approximate solution of \cite{Chimitt_2020_a}. This primarily affects step (1) of the simulation process.
    
    \item We modify the P2S basis functions to be resizable according to the camera and environmental constraints. This is done through a larger PSF training dataset which alleviates the aliasing from the previously generated set. The new P2S bases can vary from a large PSF (size $200\!\times\! 200$ or more) down to accurately modeling a delta function. This affects steps (4) and (5) of the described process.
\end{enumerate}

\subsection{Spatially varying convolution re-formulation}
\label{reformulation}
The physical meaning of a PSF is the way in which a point \emph{spreads} across the sensor plane, which we refer to as a \emph{scattering} process.
However, previous implementations of the P2S transform operate as a \emph{gathering} process. 
If the PSF is spatially invariant, the difference is trivial, equivalent to the difference between correlation and convolution. In the spatially varying case, the difference is no longer negligible. The \emph{gathering} process of previous simulators \cite{Chimitt_2020_a, Mao_2021_a, chimitt2022real} can be written as
\begin{equation}
    \boldsymbol{O} \approx \sum_{k=1}^{100} \beta_{\vx,k} \left[\boldsymbol{\psi}_k\circledast \boldsymbol{\mathcal{T}}(\boldsymbol{I}) \right]+\boldsymbol{n}.
    \label{eq: gathering_process}
\end{equation}
The \emph{scattering} process is instead written as \cite{Yanny_2022_a}:
\begin{equation}
    \boldsymbol{O} \approx \sum_{k=1}^{100} \boldsymbol{\psi}_k \circledast \left[ \beta_{\vx,k} \boldsymbol{\mathcal{T}}(\boldsymbol{I}) \right]+\boldsymbol{n}.
    \label{eq: scattering_process}
\end{equation}

While mathematically subtle, the difference is significant. Under the \emph{gathering} model, a single point source at $\vx_0$ (i.e. $\boldsymbol{\mathcal{T}}(\boldsymbol{I}) = \delta(\vx - \vx_0)$) will have the corresponding blur:
\begin{equation}
    \boldsymbol{O} \approx \sum_{k=1}^{100} \boldsymbol{\psi}_k(\vx - \vx_0) \vbeta_{\textcolor{blue}{\vx}, k}+\boldsymbol{n},
    \label{eq: blur_basis_gathering} 
\end{equation}

whereas the \emph{scattering} model \eqref{eq: gathering_process} results in
\begin{equation}
    \boldsymbol{O} \approx \sum_{k=1}^{100} \boldsymbol{\psi}_k(\vx - \vx_0) \vbeta_{\textcolor{blue}{\vx_0}, k}+\boldsymbol{n}.
    \label{eq: blur_basis}   
\end{equation}

We see \eqref{eq: blur_basis} as a shifted basis representation, whereas \eqref{eq: blur_basis_gathering} is a shifted basis with weights varying across the area of the PSF -- a mismatch to the image formation process.

\subsection{Varying $C_n^{2}$ path}
While on the surface, the problem may seem solved as described by the simulation overview. There exist some issues both at the theoretical and practical levels. The later iterations of the Zernike-based simulations \cite{Mao_2021_a, chimitt2022real} seek to rectify the practical limitations, though a key theoretical issue has remained. This leads us to introduce the two key fundamental limitations of the multi-aperture simulation:
\begin{enumerate}
    \item \textbf{Approximate solution.} Within \cite{Chimitt_2020_a}, a Taylor series is utilized to determine the correlation of the Zernike coefficients. This results in the solution only being approximate, unable to match the theoretical curves exactly as their approach utilizes a first-order Taylor approximation.
    \item \textbf{Restriction to constant $C_n^2$-paths.} Related to the Taylor series is the inability to model any turbulence beyond ground-to-ground. Furthermore, ground-to-ground situations exist for which there is a non-trivial error by the approximation, along with the potential of heat sources along the path of propagation, which would make a constant turbulence strength assumption invalid.
\end{enumerate}

These issues have been addressed by a recent analysis \cite{Chimitt_2023_a}. While it is primarily the subject of the mentioned paper, we feel it important to describe it to a sufficient level of detail here, as it is a critical improvement to the simulation quality which allows us greater accuracy in our simulations. That being said, we do not anticipate the reader who is unfamiliar with the atmospheric turbulence literature to understand the following set of equations. Therefore, we briefly present the main results for completeness and then offer an interpretation of the equations that do not require so much background.

As a wave propagates through a turbulent path, the strength of the turbulence, $C_n^2$, may vary along the propagation path. This motivates writing the strength as a function of propagation distance, $C_n^2(z)$. The new theoretical Zernike correlation result \cite{Chimitt_2023_a} allows one to write the autocorrelation of Zernike coefficients $\E[a_{\vx,i} a_{\vx',j}]$ as a function of this continuous $C_n^2$-profile:
\begin{equation}
   \E =   \mathcal{A}_{i,j}
    \int_0^L \left(\frac{L - z}{L}\right)^{5/3} C_n^2(z) f_{ij} \left( \vs(z), k_0 \right) dz
    \label{eq: main_result_continuous}
\end{equation}
where $\mathcal{A}_{i,j} = 0.00969 k^2 2^{14/3} \pi^{2/3} R^{5/3} \sqrt{(n_i + 1)(n_j + 1)}$ and $L$ is the length of propagation. The $f_{ij}$ expression is provided by \cite{Takato_1995_a}: for a displacement $\mathbf{s} =(s, \varphi)$ written in polar form, the expression in \cite{Takato_1995_a} is written as

\begin{align}
    f_{ij}(\vs, k_0) = (-1)&^{(n^+ - m^+)/2}  \Theta^{(1)}(i,j) \notag\nonumber \\ \times & I_{m^+, n_i + 1, n_j + 1}(2s, 2\pi R k_0) \notag\nonumber\\
    + (-1)&^{(n^+ +2m_i + |m^-|)/2} \Theta^{(2)}(i,j) \notag\nonumber\\ \times & I_{|m^-|, n_i + 1, n_j + 1}(2s, 2\pi R k_0),
    \label{eq: takato_expression_us_final}
\end{align}
with functions
\begin{equation}
    I_{a, b, c} (s, k_0) = \int dx \frac{J_{a}(sx) J_{b}(x) J_{c}(x)}{x(x^2 + k_0)^2},
\end{equation}
along with angular functions
\begin{equation}
    \Theta^{(1)}(i,j) = \begin{cases}
    (-1)^{j} \cos(m^+ \varphi) & h(i,j) = 1 \\
    \sin(m^+ \varphi) & h(i,j) = 2 \\
    \sqrt{2}\cos(m^+ \varphi) & h(i,j) = 3 \\
    \sqrt{2}\sin(m^+ \varphi) & h(i,j) = 4 \\
    1 & h(i,j) = 5 \\
    \end{cases}
\end{equation} and,
\begin{equation}
    \Theta^{(2)}(i,j) = \begin{cases}
    \cos(m^- \varphi) & h(i,j) = 1 \\
    \sin(m^- \varphi) & h(i,j) = 2 \\
    0 & h(i,j) = 3 \\
    0 & h(i,j) = 4 \\
    0 & h(i,j) = 5 \\
    \end{cases},
\end{equation}
contributing the angular terms and
\begin{align}
    n^\pm &= n_i \pm n_j,\\
    m^\pm &= m_i \pm m_j.
\end{align}

Though the equations which \eqref{eq: main_result_continuous} utilizes are indeed tedious to write and interpret, \eqref{eq: main_result_continuous} itself can be understood in a fairly straightforward manner. First, recall that $C_n^2(z)$ is the strength of the turbulent fluctuations. Thus, the correlation of the Zernike coefficients is a weighted summation of the turbulent distortions. The term $(L-z/L)^{5/3}$ says that turbulence \emph{closer} to the camera contributes higher strength and longer correlation length than turbulence far away from the camera. The term $f_{ij}(\cdot)$ is a result of using the Zernike polynomials -- therefore, it is simply a function that falls out of the mathematical description of them. The inner term $\vs(z)$ is a function of geometry, which ensures neighboring points have a higher correlation than points that are far apart. Finally, although $k_0$ is not so straightforward to interpret without proper background in the literature, it is related to the size of the turbulent distortions (not strength, but their geometric size).

We claim that \eqref{eq: main_result_continuous} is a significant improvement over previous results of \cite{Chimitt_2020_a}. To demonstrate this difference, we use an example as given in \cite{Chimitt_2023_a} to show that the general result \eqref{eq: main_result_continuous} contains the results of \cite{Chimitt_2020_a} as a special case. We offer some additional interpretation here to aid in understanding.

\begin{figure*}
    \centering
    \small
    \begin{tabular}{cc}
        \includegraphics[width=0.46\linewidth]{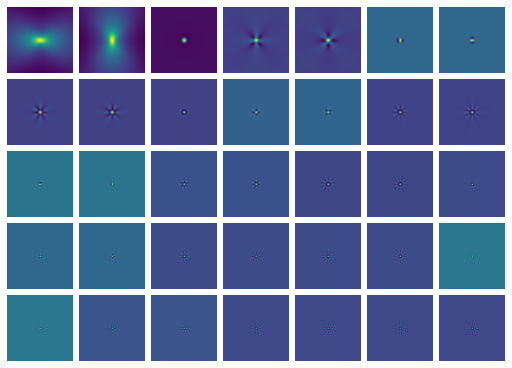} & 
        \includegraphics[width=0.46\linewidth]{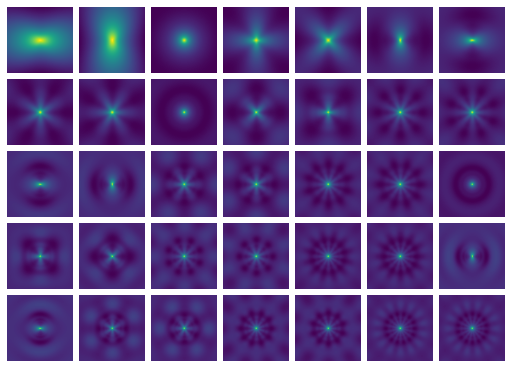} \\
        1 $C_{n}^{2}$ segment (used by \cite{Zhang_2022_a}) & 10 $C_{n}^{2}$ segments \\
        \includegraphics[width=0.46\linewidth]{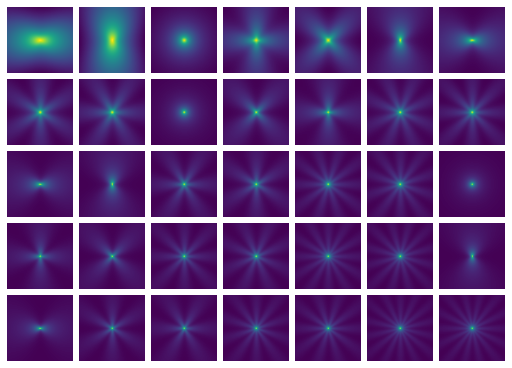} &
        \includegraphics[width=0.46\linewidth]{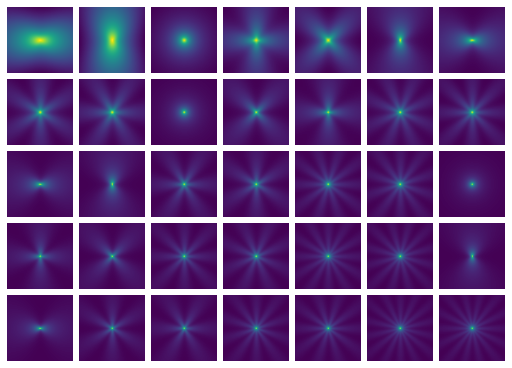} \\
        100 $C_{n}^{2}$ segments (ours) & 10000 $C_{n}^{2}$ segments
    \end{tabular}
    \caption{An instance of $\E[a_{\vx,i} a_{\vx',i}]$ from \eqref{eq: main_result_continuous} under different number of $C_{n}^{2}$ segments along the optical path. Here, we show the 2nd to 36th autocovariance functions in raster order, and brighter pixels indicate larger values. The associated parameter set is distance  = 600m, focal length = 500mm, F-number = 11, $C_{n}^{2}[z] = 5\times 10^{-14} \text{m} ^{-2/3}$ for all $z$, image size = $128\! \times\! 128$, scene width = 0.5m. From this figure, we find that the additional precision becomes negligible when we use more than 100 segments. Hence we chose 100 segments for data synthesis.}
    \label{fig: z_corr_vis}
\end{figure*}

For this example, the turbulence strength is defined to be the following
\begin{equation}
    C_n^2(z) = L C_n^2 \delta\left(z - \frac{L}{2}\right).
\end{equation}
This means the turbulence is located at the halfway point of propagation, the rest is free space. If we plug this $C_n^2(z)$ function into \eqref{eq: main_result_continuous}, we achieve the same correlation function as in \cite{Chimitt_2020_a}:
\begin{equation}
    \E[a_{\vx,i} a_{\vx',j}; 1] = \mathcal{A}_{i,j}  \left(\frac{1}{2}\right)^{5/3} L C_n^2 f_{ij}\left(\frac{ (\vx - \vx')}{D}, k_0\right).
    \label{eq: chim_chan_main_result_1}
\end{equation}
Interpreting this result means that previous Zernike-based simulations were equivalent to ``squeezing'' all of the turbulence into a single infinitesimally thin slice at the halfway point. This explains the inaccuracy by \cite{Chimitt_2020_a} as to why they cannot (i) exactly match theoretical predictions and (ii) be extended to varying $C_n^2$-profiles. Unknown to \cite{Chimitt_2020_a}, their approximation is equivalent to approximating the integral of \eqref{eq: main_result_continuous} as a single Riemann summation term.

Our approach to simulation in this paper rests on the result of \cite{Chimitt_2023_a}, which is exact. Furthermore, it does not increase time in simulation, except for a small increase in precomputation, which has been suitably optimized. We note that this precomputation happens \emph{once ever} as long as $k_0$ doesn't change (which is not too restrictive of an assumption).

To visualize the improvement in this correlation term by the number of terms used to approximate the integral \eqref{eq: main_result_continuous}, we present a visualization in \fref{fig: z_corr_vis}. This demonstrates that (i) a few additional terms contribute a great deal to the overall accuracy and (ii) an increase in terms \emph{decreases} the aliasing. The decrease in aliasing is because FFT-based generation is utilized -- any high-frequency content, which is ``blurred'' out by additional terms, may be aliased if the sample grid is not large enough spatially. (iii) Our experiments demonstrate 10-100 phase points in evaluating \eqref{eq: main_result_continuous} to be sufficient, depending on the situation.

\begin{table*}[h]
\centering
\setlength{\aboverulesep}{0pt}
\setlength{\belowrulesep}{0pt}
\footnotesize
\resizebox{0.9\textwidth}{!}{\begin{tabular}{cccccc}
\toprule[0.8pt]
Modality & Distance (m) & Focal length (m) & F-number & Scene width (m) & $C_{n}^{2} (10^{-14}\times \text{m}^{-2/3})$\\
\midrule[0.5pt]
\multirow{12}{*}{ATSyn-dynamic} & \multirow{2}{*}{[30, 100]} & \multirow{2}{*}{[0.1, 0.3]} & $ \{2.8, 4\}$ & [2, 4] & [50, 300] \\\cline{4-6}
      &   &  &  $ \{2.8, 4, 5.6\}$  & [4, 20] & [200, 1000] \\\cline{2-6}
& \multirow{2}{*}{[100, 200]} & \multirow{2}{*}{[0.2, 0.5]} & $ \{2.8, 4, 5.6\}$ & [2, 4] & [5, 50] \\\cline{4-6}
      &   &  &  $ \{2.8, 4, 5.6\}$  & [4, 20] & [20, 100] \\\cline{2-6}
& \multirow{2}{*}{[200, 400]} & \multirow{2}{*}{[0.3, 0.5]} & $\{5.6, 8\}$ & [2, 6] & [2, 30] \\\cline{4-6}
                &    &  &  $\{4, 5.6, 8\}$  & [6, 20] & [10, 40] \\\cline{2-6}
& \multirow{2}{*}{[400, 600]} & \multirow{2}{*}{[0.4, 0.75]} & $\{8, 11\}$ & [3, 7] & [1, 20] \\\cline{4-6}
                &    &  & $\{5.6, 8, 11\}$  & [7, 20] & [10, 30] \\\cline{2-6}
&\multirow{2}{*}{[600, 800]} & \multirow{2}{*}{[0.6, 0.8]} & $\{8, 11\}$ & [4, 8] & [1, 15] \\\cline{4-6}
                  &  &  &  $\{8, 11\}$  & [8, 20] & [2, 20] \\\cline{2-6}
& \multirow{2}{*}{[800, 1000]} & \multirow{2}{*}{[0.8, 1]} & $\{11, 16\}$ & [4, 8] & [0.5, 10] \\\cline{4-6}
                 &   &  &  $\{8, 11, 16\}$  & [8, 20] & [1, 20] \\
\midrule[0.5pt]
\multirow{6}{*}{ATSyn-static} & \multirow{2}{*}{[200, 400]} & \multirow{2}{*}{[1, 2]} & $\{8, 11\}$ & [0.2, 0.5] & [3, 7] \\\cline{4-6}
      &   &  & $\{5.6, 8, 11\}$  & [0.5, 1] & [6, 30] \\\cline{2-6}
& \multirow{2}{*}{[400, 600]} & \multirow{2}{*}{[1, 2.5]} & $\{8, 11, 16\}$ & [0.4, 0.8] & [2, 6] \\\cline{4-6}
                &    &  &  $\{5.6, 8, 11\}$  & [0.8, 1.5] & [6, 30] \\\cline{2-6}
& \multirow{2}{*}{[600, 800]} & \multirow{2}{*}{[1, 3]} & $\{11, 16\}$ & [0.5, 1.2] & [2, 5] \\\cline{4-6}
                &    &  &  $\{8, 11\}$  & [1.2, 2] & [5, 30] \\
\bottomrule[0.8pt]
\end{tabular}}
\caption{Parameter range, where $[a,b]$ means uniform sampling from continuous range (a, b), and $\{\}$ indicates uniform sampling from the discrete set, all rows were chosen with identical probability}
\label{table:params}
\end{table*}

\begin{table*}[h]
\centering
\small
\begin{tabular}{|c|c|c|c|c|c|}
\hline
\multirow{2}{*}{\backslashbox{Strength}{Blur}}
& \multirow{2}{*}{$k_{b}\leq 17$} & \multicolumn{3}{c|}{$19\leq k_{b} \leq 29$} & \multirow{2}{*}{$k_{b} \geq 31$} \\
\cline {3-5} & & $D/r_{0} < 2$ & $2 \leq D/r_{0} \leq 8$ & $D/r_{0} > 8$ & \\\hline
Weak & \multicolumn{2}{c|}{$\overline{d} < 0.5$} & $\overline{d} < 0.2$ & \multicolumn{2}{c|}{-} \\\hline
Medium & \multicolumn{2}{c|}{$0.5 \leq \overline{d} \leq 1$} & $0.2 \leq \overline{d} \leq 0.4$ & \multicolumn{2}{c|}{$\overline{d} \leq 0.2$} \\\hline
Strong & \multicolumn{2}{c|}{$\overline{d} > 1$} & $\overline{d} > 0.4$ & \multicolumn{2}{c|}{$\overline{d} > 0.2$} \\\hline
\end{tabular}
\caption{Turbulence strength criterion in ATSyn-dynamic, the value of $k_{b}$ is odd.}
\label{table:strength}
\end{table*}

\subsection{New P2S kernels}
In an optical simulation, careful consideration of the various sample spacings is critical for achieving high accuracy. Previous multi-aperture simulations have made some progress in this direction. However, their approach is limited in many ways. The reason for this reduces to the fact that their kernels $\vpsi_i$ may not be easily resized. This hurts the accuracy of the simulation by causing mismatches in sampling and limits the model's generalizability. 

The P2S kernels implemented in this paper are (i) resizable and (ii) chosen to match the sampling parameters of the scene. The core solution is (i), with (ii) being an important consequence of this correction. The main limitation in the P2S bases is their initial size of $33\!\times\!33$. This causes the bases too often to be aliased significantly upon resizing. To address this issue, we have increased the resolution of the PSF dictionary, resulting in the basis functions being of size $67\!\times\!67$. Additionally, the dictionary is $20\!\times$ larger than \cite{Mao_2021_a}, aiding in the eigenfunctions being well-behaved. The dictionary is generated with turbulence strength $D/r_{0} = [0.1, 12]$, representing various turbulent conditions. Through our testing, we have observed we can match PSFs from a delta function up to the challenging cases of $6 \leq D/r0 \leq 12$.

With these modifications, we have observed no notable aliasing when resizing the PSF basis functions. This allows us to resize the bases to match the sampling specified in the simulation parameters. This is done by a tuning step that operates in the following way:
\begin{enumerate}
    \item The basis is used to represent the diffraction kernel offline. We can compute the full width at half maximum (FWHM) in pixels for the basis $N_d$. This step is done once and hard-coded into the simulation.
    \item Given the specified image size and camera parameters, the diffraction kernel FWHM can be computed in meters and converted to pixels $N_0$. This is done for every new set of parameters.
    \item The basis is resized by $N_0/N_d$, making the FWHM of the diffraction kernel coincide with the theoretically predicted value.
\end{enumerate}
Through this process, we can correctly incorporate the sampling of the imaging system and scene into the basis representation. In addition, we optionally incorporate PSF basis size scaling by $D/r_0$. We have observed that this gives us additional turbulence blur not captured in the above PSF resizing scheme.

\subsection{Temporal correlation}
Real-world turbulence is temporally correlated because the dynamics of the atmosphere is a continuous process. Therefore, accurately simulating a video will require the degradation to be spatiotemporally correlated. We disentangled the spatial and temporal correlation and injected temporal correlation into the simulation process by correlating the initial random seed in the simulation. We use an $AR(1)$ process to generate the initial seed at the first stage. This allows for the random seed $\boldsymbol{n}_{t}$ at time $t$, which is then used to form the distortion and blur random fields, to be related to the previous realization by
\begin{equation}
\boldsymbol{n}_{t} = \alpha \boldsymbol{n}_{t-1} + \sqrt{1-\alpha^{2}}\boldsymbol{\epsilon}_{t}
\end{equation} 
The term $\alpha$ is the temporal correlation ratio and $\boldsymbol{\epsilon}_{t} \sim \mathcal{N}(0, \boldsymbol{I})$.

\section{ATSyn Dataset}
\label{dataset}
The ATSyn dataset has two subsets: \emph{ATSyn-dynamic} and \emph{ATSyn-static}. The objective of the static scene turbulence mitigation task is to restore a single common ground truth from a sequence of degraded frames, which has been extensively explored in classical turbulence mitigation literature. On the other hand, the dynamic scene turbulence mitigation task aims to restore each video frame where the object or scene is in motion, presenting a significantly greater challenge for conventional methods. As stated in the main paper, the \emph{ATSyn-dynamic} contains 5447 groups of turbulence-affected videos, the $\boldsymbol{\mathcal{T}}$-only videos and ground truth videos. Among all 5447 groups, 4350 are for training, and 1097 are for validation. Frame-wise, we have 1816375 frame groups for training. We use the first 120 frames in each testing video during testing if the original testing video has more than 120 frames. On the other hand, the ATSyn-static subset contains 3000 groups of image sequences, each consisting of 50 turbulence-affected frames, 50 $\boldsymbol{\mathcal{T}}$-only frames, and a corresponding ground truth image. Out of these 3000 groups, 2000 are designated for training, while 1000 are set aside for validation. Thanks to the efficiency of our simulator, the entire synthesis process can be completed within seven days using a single RTX 2080Ti GPU or 42 hours using a single NVIDIA A100 GPU.

\subsection{Parameter selection details}
\label{param_select}
Using the simulation method in Section 1, we can synthesize long-range atmospheric turbulence effects at various physical and camera parameters. These parameters include distance, the field of view (FOV) represented by scene width, turbulence profile indicator $C_{n}^{2}$, focal length, and F-number of the camera. The detailed parameter ranges are shown in Table \ref{table:params}. When setting the parameters, we first select the distance, FOV, focal length, and f-number with parameters ranging from a standard camera and lens to an astronomical telescope. We then choose the $C_{n}^{2}$ range to set the turbulence effect to be neither too strong nor weak. The temporal correlation was sampled from $0.2\! \sim\! 0.9$ in the \emph{ATSyn-static} and $0.3\! \sim\! 0.95$ in the \emph{ATSyn-dynamic}.

\subsection{Turbulence strength}
We classify the turbulence strength into multiple levels to study how turbulence mitigation networks perform under different conditions. For the ATSyn-dynamic dataset, we select three levels. Although our parameters are carefully chosen, the relationship between turbulence strength and parameters is highly nonlinear. We, therefore, determined the turbulence strength based on the actual degradation of the image. Turbulence degradation consists of the pixel displacement and blur effect. The average pixel displacement (denoted by $\overline{d}$) can measure the former. The latter can be indicated by the size of the blur kernel basis (denoted by $k_{b}$) and the turbulence strength $D/r_{0}$. The size of the blur kernel basis is related to, though not proportional to, $D/r_{0}$; the blur kernel size is also affected by the image resolution, distance, and field of view. It is possible that the same blur kernel basis yields different blur effects under different $D/r_{0}$ or that the same $D/r_{0}$ is associated with different blur sizes because the resolution of the blur kernel varies. Therefore, we need to consider both the size of the basis and $D/r_{0}$. The detailed classification criterion is listed in Table \ref{table:strength}.

We use 4500 clean input videos to generate the dataset, partitioned into three groups with 1500 videos per partition. For each video, we run the parameter generator in Section \ref{param_select} to produce random turbulence parameters and synthesize a single sample frame. The turbulence strength can be determined from this instance according to Table \ref{table:strength}. We synthesize the entire video if the associated turbulence strength set is not full, or we abandon the set of parameters and randomly produce another set and repeat the steps above until the video is accepted by one turbulence strength set or all videos are synthesized.

\end{appendices}

{
    \small
    \bibliographystyle{ieeenat_fullname}
    \bibliography{main}
}


\end{document}